\documentclass[letterpaper,12pt]{JHEP3}
\usepackage{graphicx}
\usepackage{amsmath}
\usepackage{amsfonts}
\usepackage{amssymb}
\usepackage{feynmp}
\def\beq{\begin{equation}}
\def\eeq{\end{equation}}
\def\bea{\begin{eqnarray}}
\def\eea{\end{eqnarray}}
\def\lsim{\mathrel{\raise.3ex\hbox{$<$\kern-.75em\lower1ex\hbox{$\sim$}}}}
\def\gsim{\mathrel{\raise.3ex\hbox{$>$\kern-.75em\lower1ex\hbox{$\sim$}}}}
\def\ifmath#1{\relax\ifmmode #1\else $#1$\fi}
\def\half{\ifmath{{\textstyle{1 \over 2}}}}

\def\quarter{\ifmath{{\textstyle{1 \over 4}}}}

\def\fbi{~{\mbox{fb}^{-1}}}
\def\fb{~{\mbox{fb}}}
\def\pb{~{\mbox{pb}}}
\def\br{BR}
\def\gev{~{\mbox{GeV}}}

\def\calm{\mathcal{M}}
\def\to{\rightarrow}

\def\ptmiss{p_T^{\rm miss}}
\def\etmiss{E_T^{\rm miss}}

\def\wtil{\widetilde}
\def\nn{\nonumber}
\def\half{{1\over 2}}
\def\quarter{{1\over 4}}

\def\calm{{\cal M}}
\def\del{\delta}

\def\sig{\sigma}
\def\anti{\overline}

    \def\fillboxx#1#2{\hbox to #1{\vbox to #2{\vfil}\hfil}   }

\def\gev{~{\rm GeV}}

\def\tanb{\tan\beta}

\def\cnone{\wt\chi^0_1}

\def\cntwo{\wt\chi^0_2}

\def\mcnone{m_{\cnone}}
\def\mcntwo{m_{\cntwo}}

\def\wt{\widetilde}
\def\anti{\overline}

\def\stauone{\wt \tau_1}
\def\mstauone{m_{\stauone}}

\def\gl{\wt g}
\def\mgl{m_{\gl}}

\def\sbot{\wt b}
\def\msbot{m_{\sbot}}

\def\slep{\wt \ell}

\def\mslep{m_{\slep}}

\def\slepr{\wt \ell_R}
\def\mslepr{m_{\slepr}}

\newcommand{ \slashchar }[1]{\setbox0=\hbox{$#1$}   
   \dimen0=\wd0                                     
   \setbox1=\hbox{/} \dimen1=\wd1                   
   \ifdim\dimen0>\dimen1                            
      \rlap{\hbox to \dimen0{\hfil/\hfil}}          
      #1                                            
   \else                                            
      \rlap{\hbox to \dimen1{\hfil$#1$\hfil}}       
      /                                             
   \fi}     
\title{Mass Determination in SUSY-like Events with Missing Energy}

\author{Hsin-Chia Cheng, John F. Gunion, Zhenyu Han, Guido Marandella,
and Bob McElrath\\ Department of Physics, University of California,
Davis, CA 95616\\ E-mail:
\email{cheng@physics.ucdavis.edu, jfgucd@physics.ucdavis.edu,
zhenyuhan@physics.ucdavis.edu, maran@physics.ucdavis.edu,
mcelrath@physics.ucdavis.edu}}

\abstract{ We describe a kinematic method which is capable of
  determining the overall mass scale in SUSY-like events at a hadron
  collider with two missing (dark matter) particles.  We focus on the kinematic
  topology in which a pair of identical particles is produced with
  each decaying to two leptons and an invisible particle
  (schematically, $pp\to YY+jets$ followed by each $Y$ decaying via
  $Y\to \ell X\to \ell\ell'N$ where $N$ is invisible). This topology
  arises in many SUSY processes such as squark and gluino production
  and decay, not to mention $t\anti t$ di-lepton decays.  In the
  example where the final state leptons are all muons, our errors on
  the masses of the particles $Y$, $X$ and $N$ in the decay chain
  range from 4 GeV for 2000 events after cuts to 13 GeV for 400 events
  after cuts. Errors for mass differences are much smaller. Our
  ability to determine masses comes from considering all the kinematic
  information in the event, including the missing momentum, in
  conjunction with the quadratic constraints that arise from the $Y$,
  $X$ and $N$ mass-shell conditions. Realistic missing momentum and
  lepton momenta uncertainties are included in the analysis.}

\begin{document}

\section{Introduction}
\label{sec:introduction}

As the Large Hadron Collider (LHC) is near completion, we will soon be
able to fully explore TeV scale physics. 
Because of the naturalness problem for the Higgs boson in the
context of the Standard Model (SM), it is strongly believed that new
physics beyond the SM should appear at or below the TeV
scale. There are many
possible candidates for TeV-scale physics beyond the Standard Model, giving rise
to various experimental signatures at the LHC. If some new signal is
discovered, it is vital to determine the masses and spins of the new
particles in order to fully reconstruct the picture of the TeV scale.

Some new physics will be easily identified. For example, if there is a
$Z'$ gauge boson accessible at the LHC, one can easily find it by
looking for the resonance in the invariant mass distributions of its
decay products, {\it e.g.,} a pair of leptons or jets. In general, if
the decays of a new particle involve only visible particles, one can
search for it by looking for a bump in various invariant mass
combinations of the visible particles and the location of the bump
determines the mass of the new particle. On the other hand, if the
decays of a new particle always contain one or more invisible
particles, the search for the new particle becomes more complicated,
as there is no ``bump'' to look for. In order to detect new physics in
such a case, it is necessary to understand the SM
backgrounds very well and to then look for excesses above them.
Determining the masses of the new particles will also be challenging
since we cannot directly measure the energy carried away by the
invisible particles. Absent good mass determinations, it will be
difficult to reconstruct a full picture of the TeV scale even after
new physics is discovered.

A scenario with missing particles is highly motivated for TeV scale
physics, independent of the hierarchy problem.
If we assume that dark matter  is the thermal relic of some
weakly interacting massive particles (WIMPs) left from the Big Bang,
then the right amount of dark matter in the universe is obtained for a
WIMP mass in the 0.1--1~TeV range under the assumption that the
electroweak sector mediates the dark matter---SM
interaction.  The dark matter particle must be electrically neutral and
stable on cosmological time scales. If it is produced at a collider, 
because it is weakly interacting it
will escape the detector without being detected, giving missing energy
signals. In order for the dark matter particle to be stable, it is
likely that there is a new symmetry under which the dark matter particle
transforms but all SM particles are neutral, thereby
preventing decay of the dark matter particle to SM particles.

LEP has indirectly tested physics beyond the SM. The electroweak
precision fit and 4-Fermi contact interaction constraints exclude new
particles with masses below $\mathcal{O}({\rm TeV})$ if they are exchanged at tree
level, unless their coupling to the SM fermions is suppressed. If
there is a symmetry under which the new particles are odd and the SM
particles are even, then the new particles can only contribute to the
electroweak observables at the loop level. In this case, the bound on the mass of
the new particles decreases by about a loop factor, $m \to m/4 \pi$,
making the existence of new particles with masses of order a few
hundreds of GeV compatible with the data.  The message coming from the
LEP data is that, if there is any new physics responsible for
stabilizing the electroweak scale, it is very likely to possess such a
new symmetry. Thus, the cosmological evidence for dark matter together
with the LEP data provide very strong motivation for new particles at
or below the TeV scale that are pair produced rather than singly
produced.

Almost all the models with dark matter candidates contain
additional particles charged under the new symmetry. At a collider,
these new particles must also be pair-produced, and if they are
heavier than the dark matter particle, they will cascade decay down to
it.  In many cases, this cascade radiates SM particles in a series of
$A\to Bc$, $1 \to 2$ decays, in which $A$ and $B$ are new physics
particles while $c$ is a SM particle.  (In some cases, phase space
restrictions force one of the new particles off-shell and $A\to B^*
c\to C d c$, $1\to 3$ decays are relevant.) Since the final step in
the chain will yield a dark matter particle, the typical collider
signals for such a scenario will be jets and/or leptons plus missing energy.

Supersymmetry (SUSY) is the most popular candidate for physics beyond
the SM and belongs to the above category of models. In
SUSY, $R$-parity conservation implies that the Lightest Supersymmetric
Particle (LSP) is stable. In most supersymmetric models the LSP is the
lightest neutralino, which is a good dark matter candidate. It appears
at the end of every supersymmetric particle decay chain and escapes
the detector. All supersymmetric particles are produced in pairs,
resulting in at least two missing particles in each event.

Other theories of TeV-scale physics with dark matter candidates have been
recently proposed. They have experimental signatures very similar to SUSY:
{\it i.e.} multiple leptons and/or jets plus missing energy. For instance,
Universal Extra Dimensions (UEDs)~\cite{ued,Cheng:2001an}, little Higgs
theories with $T$-parity (LHT)~\cite{lht}, and warped extra dimensions
with a $Z_3$ parity~\cite{Agashe:2004ci} belong to this category of
models.  Being able to reconstruct events with missing energy is
thus an important first step to distinguish various scenarios and
establish the underlying theory.

Of particular importance will be the determination of the absolute
masses of the new particles, including the dark matter particle.
First, these masses are needed in order to determine the underlying
theory. For example, in the case of SUSY, accurate particle masses are
needed to determine the SUSY model parameters, in particular the
low-scale soft-SUSY-breaking parameters.  These in turn can be evolved
to the unification scale (under various assumptions, such as no
intermediate-scale matter) to see if any of the proposed GUT-scale
model patterns emerge. The accuracy required at the GUT-scale after
evolution implies that low-scale masses need to be determined with
accuracies of order a few GeV. Second, the mass of the dark matter particle,
and the masses of any other particles with which it can coannihilate,
need to be determined in order to be able to compute the dark matter relic
density in the context of a given model. Studies \cite{Baltz:2006fm}
suggest that the required accuracy is of order a few GeV. A very
important question is then whether or not the LHC can achieve such
accuracy or will it be necessary to wait for threshold scan data from
the ILC.  One goal of this paper will be to find techniques for
determining the dark matter particle mass at the LHC with an accuracy that 
is sufficient for a reasonably precise computation of the relic density.

Most of the SUSY studies carried out thus far have relied on long
decay chains of super-particles which produce many jets and/or leptons and
large missing energy. Several kinematic variables have been proposed 
as estimators of the super-particle mass scale, such as $\not\!\!E_T$,
$H_T$, $M_{eff}$~\cite{Hinchliffe:1996iu}, and
$M_{T_2}$~\cite{cambridge}.  However, these
variables measure the mass differences between the super-particles, but
not the overall mass scale.

One possible means for determining the overall mass scale is to
employ the total cross section. However, the total cross section is
very model dependent: it depends on the couplings, the species of the
particles being produced, {\it e.g.,} fermions or bosons, as well as
the branching fractions of the decays involved in the process.  One
needs to have already determined the spins and branching ratios for
this to be reliable, a task that is difficult or impossible at the LHC
without an ability to determine the four-momenta of all the particles
involved in the process. To fully test a potential model, we must
first determine the masses of the produced particles using only
kinematic information.  Once the masses are known, there are many
chain decay configurations for which it will be 
possible to use these masses to determine the four-momenta of all the particles on an
event-by-event basis. The four-momenta can then be employed in
computing the matrix element squared for different possible spin
assignments.  In this way, a spin determination may be possible, and
then the cross section information can be used to distinguish
different models.

\begin{figure}
       \begin{center}
\includegraphics[width=4in]{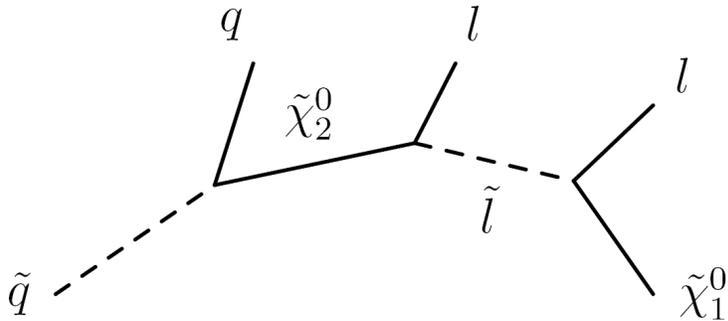}
     \caption{\label{fig:chain_decay}A decay chain in SUSY.}
      \end{center}
\end{figure}

In recent years there have been numerous studies of how to measure the
super-partner masses just based on kinematic
information~\cite{Hinchliffe:1996iu,Bachacou:1999zb,Allanach:2000kt,Gjelsten:2004ki,Weiglein:2004hn,Kawagoe:2004rz,Allanach:2004ub,Lester:2005je,Arkani-Hamed:2005px,Butterworth:2007ke}.
These studies rely on long decay chains of super-particles, usually
requiring 3 or more visible particles in the decay chain in order to
have enough invariant mass combinations of the visible particles. One
can then examine the kinematic edges of the distributions of the various
invariant mass combinations, and obtain the masses from the relations
between the end points of the distributions and the masses. Many of
these studies use the decay chain $\tilde{q} \to \tilde{\chi}^0_2 q \to
\tilde{\ell}\ell q \to \tilde{\chi}^0_1 \ell \ell q$
(Fig.~\ref{fig:chain_decay}) that occurs for the benchmark point
SPS1a~\cite{Allanach:2002nj}, for which $\mcnone\sim 97\gev$,
$\mslep\sim 143\gev$, $\mcntwo\sim 180\gev$, $\msbot\sim 570\gev$ and
$\mgl\sim 610\gev$, see Appendix B.  The kinematic endpoints of the
invariant mass distributions, $m_{\ell\ell}$, $m_{q\ell\ell}$, $m_{q\ell ({\rm
high})}$, and $m_{q\ell ({\rm low})}$\footnote{High and low represent
the largest and the smallest values of $m_{q\ell}$,
respectively --- these masses are employed since it is
not possible to determine 
the order in which the observed leptons appear in the chain
decay.}, depend on the masses of the super-particles in the decay chain
through some complicated
relations~\cite{Allanach:2000kt,Gjelsten:2004ki,Miller:2005zp}. If the end points of
these distributions can be accurately determined from the experimental
data, we can invert the relations to obtain the masses of the
super-particles. 

For the decay chain of Fig.~\ref{fig:chain_decay} and the specific
model points studied, this approach can give a reasonable determination of
the masses of the super-particles, but there is room for improvement.
In some of the studies, it is only mass differences that are well
determined whereas the overall mass scale is rather uncertain.  For
one of the mass points studied in~\cite{Gjelsten:2004ki,Miller:2005zp}
(labeled $\alpha$ in \cite{Gjelsten:2004ki}), a very large number of
events is employed and the overall mass scale uncertainty is reduced
to two discrete choices, one corresponding to the correct solution
(with rms error relative to the central value of order 4 GeV) and the
other (somewhat less probable value) shifted by about 10 GeV.  For the
mass choices labeled $\beta$, for which the event rate is lower, there
are a number of discrete solutions and each one has fairly large rms
error for the absolute mass scale. However, it should be noted that in
reducing the solutions to a number of discrete choices, not only were
the locations of the kinematic edges employed, but also the shapes of
the distributions of the mass combinations were employed.  These
latter shapes depend upon their choice of model being correct.  It is
possible that without this information there would have been a
significant continuous range of overall possible mass scale.

Another mass determination method is that proposed by Kawagoe, Nojiri,
and Polesello~\cite{Kawagoe:2004rz}. Their method relies on an even
longer decay chain, $\tilde{g} \to \tilde{b} b_2 \to \tilde{\chi}^0_2
b_1 b_2 \to \tilde{\ell} b_1 b_2 \ell_2 \to \tilde{\chi}^0_1 b_1 b_2
\ell_1 \ell_2$. There are five mass shell conditions and for each
event there are the four unknowns due to the unobservable 4-momentum
of the $\tilde{\chi}^0_1$.  In principle, before introducing
combinatorics and experimental resolution, one can then find a
discrete set of solutions in the space of the 5 on-shell masses as
intersections of the constraints coming from just five events. In
practice, combinatorics and resolution complicate
the picture. In their actual analysis, they only fitted the gluino and
the sbottom masses {\it with the assumption that the masses of
$\tilde{\chi}^0_2$, $\tilde{\ell}$, and $\tilde{\chi}^0_1$ are already
known}. For the standard SPS1a point, they achieved accuracies for $m_{\tilde g}$ and $m_{\tilde
  b}$ of order a few GeV, but with central values systematically
shifted (upward) by about 4 GeV.  In a follow up study
\cite{Allanach:2004ub}, Lester discusses a procedure for using all 5
on-shell mass constraints. For a relatively small number of
events and without considering the combinatorics associated with 
the presence of two chains, he finds a  $17\%$ error
in the determination of $\mcnone$.

In addition to the above studies, a series of contributions concerning
mass determination appeared in \cite{Weiglein:2004hn}. These latter
studies focused on the SPS1a point and again employed the kinematic
edges of the various reconstructable mass distributions in the
$\tilde{g} \to \tilde{b} b_2 \to \tilde{\chi}^0_2 b_1 b_2 \to
\tilde{\ell} b_1 b_2 \ell_2 \to \tilde{\chi}^0_1 b_1 b_2 \ell_1
\ell_2$ decay chain to determine the underlying sparticle masses.
Experimental resolutions for the jets, leptons and missing energy
based on ATLAS detector simulations were employed. The resulting final
errors for LHC/ATLAS are quoted in Table 5.1.4
of~\cite{Weiglein:2004hn}, assuming an integrated luminosity of
$300\fbi$ and after using both $\wtil e$ and $\wtil \mu$ intermediate
resonances (assuming they are degenerate in mass).  We have since
verified with several ATLAS members that the quoted errors do indeed
correspond to $\pm 1\sigma$ errors \cite{dirk}. The tabulated errors
for $\mcnone,\mslep,\mcntwo$ are all of order $5\gev$, while those for
$\msbot$ and $\mgl$ are of order $8\gev$.

In all of the studies referenced above, the methods employed required
at least three visible particles in the decay chain, and, in the last
cases above, four visible particles (two $b$'s and two $\ell$'s).  We
will study the seemingly much more difficult case in which we make use
of only the last two visible particles in each decay chain. (For
example, the subcase of Fig.~\ref{fig:chain_decay} in which only the
$\cntwo \to \ell\ell\cnone$ portion of each decay chain is employed.)
In this case, if only the isolated chain-decays are analyzed, the one
invariant mass combination that can be computed from the two visible
4-momenta does not contain enough information to determine the three
masses ($\mcntwo$, $\mslep$ and $\mcnone$) involved in the decay
chain. Thus, we pursue an alternative approach which employs both
decay chains in the event at once. For the SPS1a point, our method
allows a determination of the masses $\mcntwo$, $\mslep$ and $\mcnone$
with an accuracy of $\sim \pm 5\gev$ after including both $\wtil e$
and $\wtil \mu$ intermediate slepton states (again, taken to be
degenerate in mass), assuming $L=300\fbi$ and
adopting the ATLAS expectations for the resolutions for lepton
momentum and missing momentum measurements.  (These resolutions affect
the determination of the crucial transverse momentum of the
$4\ell+2\cntwo$ system.  In particular, by looking at only the
leptonic part of the decay chains we can avoid considering individual
jet momenta, and therefore we are less sensitive to imprecise
measurements for the individual jets.) In short, using only the leptons
in the final state, we obtain an accuracy that is very comparable to
the $\sim \pm 5\gev$ found in the LHC/ATLAS study referenced above for
the same luminosity and very similar detector simulation.

In the above single-chain decay approaches, it is implicitly assumed
that appropriate cuts {\it etc.} have been employed so that both decay
chains in each event involve the same decaying resonances, all the way
back to the $\gl$. In our approach it is unnecessary to know exactly
what resonances appear prior to the $\cntwo$'s in the two decay
chains. Thus, some of the $\cntwo$ pair events could come from direct
$\wtil q$ production and some indirectly from $\wtil g$ production
followed by $\gl\to q \wtil q$ decay.  We also do not need to tag the
$b$ quarks. We only need to measure to determine the transverse
momentum of the $\cntwo\cntwo$ pair using the measured lepton momenta
and the measured missing momentum.
Nonetheless, we do need to isolate a sample of events
dominated by two final $\cntwo\to \ell \wtil \ell\to
\ell\ell\cnone$ decays. (Of course, it
is interesting to go beyond this assumption, but we will not do so in
this paper.)  The key to mass determination using the more limited
information we employ is to consider the whole event at once and look
not for edges in masses reconstructed from visible momenta but for
sharp transitions in the number of events consistent with the assumed
topology after an appropriate reconstruction procedure. Further,
as noted later, if the events we isolate do not correspond to a
$\cntwo\cntwo$ pair decaying in the manner assumed, then our
procedure will yield smooth distributions in the number of
reconstructed events, as opposed to the sharp transitions predicted if we
have indeed isolated an enriched $\cntwo\cntwo$-pair sample with
decays as presumed.

\begin{figure}
\begin{center}
\includegraphics[scale=0.5,angle=0]{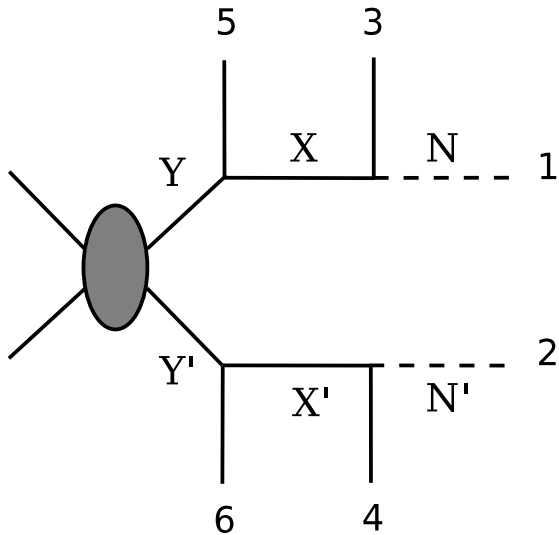}
\end{center}
\caption{\label{fig:topology}
 The event topology we consider.}
\end{figure}

Beginning with the general topology illustrated in
Fig.~\ref{fig:topology}, we employ the information coming from
correlations between the two decay chains in the same event, and the
missing momentum measurement.  This is evident from some simple
constraint counting. Each event satisfying the topology of
Fig.~\ref{fig:topology} has
the two invisible 4-momenta of the $N$ and $N'$. The sum of the
transverse momenta of $N$ and $N'$ is, however, constrained to equal
the negative of the sum of the transverse momenta of the visible
particles, leaving us with 6 unknowns for each event, subject to 6
on-shell mass constraints.  Under the assumption that $m_Y=m_{Y'}$,
$m_X=m_{X'}$ and $m_N=m_{N'}$, we are left with the three unknown
masses, $m_Y$, $m_X$ and $m_N$.  Every event will be compatible with a
certain region in the 3-dimensional $\{m_Y,m_X,m_N\}$ space.
Combining several events will shrink this region.  We will show that
before the inclusion of combinatorics and resolution effects the
actual values of the masses lie at the end point of such a region.
Mass determination after including combinatoric and resolution effects
requires an examination of how the number of events consistent with
given mass choices changes as the masses are shifted.

In our approach, we find that it is important to not focus on the
individual invariant mass distributions, as this would not utilize all
the information contained in the data. Instead, we examine the events
from a more global point of view and try to use all the kinematic
information contained in the data to determine the masses of the
particles involved.\footnote{We note that the fact that each event
  defines a region in mass space was also the case for the $Z\to Y \to
  X \to N$ one-sided chain situation outlined earlier (except for the
  mass space being 4-dimensional).  An interesting question is whether
  our more general approach would determine the absolute mass scale in
  the one-sided case, as opposed to just mass differences.  A detailed
  study is required.} In the case where we have of order 2000 events
available after cuts, and after including combinatorics and
resolutions for missing momentum and lepton momentum measurements according to ATLAS
detector expectations, we achieve rms accuracies on $m_Y$, $m_X$ and $m_N$
of order 4 GeV, with a small
systematic shift that can be easily corrected for. This assumes a case
with significant separation between the three
masses. This result is fairly stable when backgrounds are included so
long as $S/B\gsim 2$. This number of events and resulting error apply
in particular to the SPS1a point assuming integrated luminosity of
$L=300\fbi$ and use of all $\slep =\wtil e$ or $\wtil \mu$ channels.

The organization of the paper is as follows.  In
Sec.~\ref{sec:topology}, we give a detailed exposition regarding
solving the topology of Fig.~\ref{fig:topology}.  In
Sec.~\ref{sec:no_resolution}, we demonstrate how the masses of the
$Y$, $X$ and $N$ particles in Fig.~\ref{fig:topology} can be very
precisely determined after a reasonable number of events ({\it e.g.}
500) if there are no uncertainties associated with combinatorics or
with particle and missing momentum measurement resolutions. In
Sec.~\ref{sec:with_resolution}, we develop the very crucial strategies
for dealing with the realistic situation where combinatorics,
resolutions and backgrounds are included. Sec.~\ref{sub:rescomb}
focuses on the effects of combinatorics and finite resolutions
for the lepton and measurement missing momentum measurements. 
If of order 2000 events are present
after cuts, we still find good accuracies for not only mass
differences, but also for the absolute mass scale, using only the
kinematic information contained in the available events.  In
Sec.~\ref{sub:backgrounds}, we 
discuss the effects of having background events mixed with the signal
events. In Sec.~\ref{sec:otherpoints}, we discuss two alternative scenarios:
one with very different $m_Y-m_X$ compared to the first point
analyzed, and one with $m_N\sim 0$. In Sec.~\ref{sub:sps1a}, we
consider in detail the
SPS1a mSUGRA point.  We summarize and present additional discussion in
Sec.~\ref{sec:discussion}.

\section{Topology of events with missing energy}
\label{sec:topology}

We study the collider events with topology shown in
Fig.~\ref{fig:topology}.  
A hard hadronic collision produces two identical or mirrored chains.
Each decay chain gives two visible particles and one missing particle.
It will be convenient to label the 6 final outgoing
particles from 1--6, with $N=1$, $N'=2$, visible particles $3$ and $5$ emitted
from the $Y$ chain and visible particles $4$ and $6$ emitted from the $Y'$
chain.  There are many processes which have this topology. For example,
$t\bar{t}$ production, with $t$ decaying to $bW$ and $W$ decaying
leptonically to $\ell \nu$, is exactly described by this topology, so
it can be studied with our method, except that we already know that
neutrinos are (to a good approximation) massless. There are also many
SUSY or other beyond the SM processes which can be described by this
topology, {\it e.g.,} second neutralino pair production
$\tilde{\chi}^0_2 \tilde{\chi}^0_2$ (through $t$-channel squark
exchange) with $\tilde{\chi}^0_2 \to \ell \tilde{\ell}$ and then
$\tilde{\ell} \to \ell \tilde{\chi}^0_1$, producing 4 visible charged
leptons and 2 missing particles. As already noted, we require that the
masses of the corresponding particles in the two chains be the same.
They can be the same particle or one can be the anti-particle of the
other.  Or, they can even be different particles whose masses are
believed to be approximately equal ({\it e.g.,} squarks or sleptons of
the first two generations).  The visible particles do not need to be
stable as long as we can see all their decay products and reconstruct
their 4-momenta.  The event can involve more particles (such as
hadronic ISR/FSR or parent particles such as squarks and gluinos
decaying within the gray blob on Fig.~\ref{fig:topology}) as long as
none of the additional particles lead to missing momentum.  For
example, the 4 leptons plus missing energy event from the decays of a
pair of second neutralinos can be part of the longer decay chains from
squark pair production, as occurs  for the  SPS1a chain decay.

It is instructive to analyze the unknowns in this topology in a more
detailed manner than given in the introduction. In particular, we can make a
distinction between {\it kinematic unknowns} --- those in which phase space
is differential --- and {\it parametric unknowns} --- Lagrangian parameters
or otherwise non-kinematic unknowns on which the cross section has some
functional dependence.  For instance in the Breit-Wigner propagator
$[(q^2-M^2)^2+M^2\Gamma^2/4]^{-1}$, $q$ is kinematic while $M$ and
$\Gamma$ are parametric.  Masses, including those of missing particles,
are parametric unknowns (phase space $d^3p/2 E$ is not differential in
them).  Any function of an event's missing 3-momenta and already-known
parameters is a kinematic unknown.

Each event with the topology of Fig.~\ref{fig:topology} has eight
kinematic unknowns: $\vec{p}_N, 
\vec{p}_{N^\prime}$ and the initial state $E$ and $p_z$, where we are
assuming the parameters $m_N$ and $m_{N'}$ are fixed. Total 4-momentum
conservation reduces this to four kinematic unknowns.  In the narrow
width approximation, the mass combinations constructed from a
combination of visible momenta and invisible momenta (which we place
in the class of kinematic unknowns), such as $m_{13}^2\equiv
(p_1+p_3)^2$, are equal to the corresponding parametric unknowns, such
as $m_{X}^2$, and we can trade them for their corresponding
parameters.  Therefore, in the narrow width approximation, a single
event is described by a volume in the six dimensional parameter space
$\{m_Y,m_{Y^\prime},m_X,m_{X^\prime},m_N,m_{N^\prime}\}$.

If the two chains are identical or mirrors of one another and if we
use the narrow width approximation, we can impose 3 more relations,
$m_Y=m_{Y'}$, $m_X= m_{X'}$ and $m_N= m_{N'}$, which reduces the
independent unknown parameters to three. As a result, if we know the
three masses $m_Y$, $m_X$, and $m_N$ then (up to discrete ambiguities
associated with multiple solutions to a quartic equation) we can solve
for all the unknown momenta, using the measured visible momenta, and
vice versa. The procedure is described in more detail in Appendix
~\ref{sec:relations}.

If the masses are not known, we must assume values for the three
masses $m_Y$, $m_X$, and $m_N$. Given a fixed $\calm=\{m_Y,m_X,m_N\}$
choice, for each event we obtain a quartic equation (for the energy of
one of the invisible particles) with coefficients depending on the
assumed masses, $\calm$, and visible momenta. It can have 0 to 4 real
solutions for the invisible energy, depending on the coefficients, and
each solution fully determines associated 4-momenta for both invisible
particles.

Any solution with real and physically
acceptable invisible 4-momenta corresponds to a choice for $m_Y$,
$m_X$, and $m_N$ that is consistent with that particular event.  The
points in $\calm=\{m_Y,m_X,m_N\}$ parameter space that yield real solutions
are not discrete; instead, each event defines a region in the
three-dimensional mass space corresponding to a volume of real
solutions.  The region in the mass space consistent with all events,
the `allowed' region,
will shrink as we consider more and more events.  However, even for
many events the allowed
region remains three-dimensional and does not shrink to a
point.  We need to find techniques that allow us to identify the
correct mass point given a volume in mass space consistent with a set
of events.

\section{Idealized case: Perfect resolution, no combinatorics and no background}
\label{sec:no_resolution}

In order to understand how the mass information is contained in the
kinematics, we start with the ideal case in which all visible momenta
are assumed to be measured exactly and we associate each lepton with
the correct chain and position in the chain ({\it i.e.}, we neglect
resolution effects and combinatorics).  For illustration, we have
generated a sample of 500 events of $\tilde q_L\tilde q_L$ production,
with each $\tilde q_L$ decaying according to
Fig.~\ref{fig:chain_decay}, with $Y=Y'=\cntwo$, $X=X'=\wt \mu_R$,
$N=N'=\cnone$, and 3, 4, 5, 6 all being $\mu$'s of various signs.  We
generated our events using SHERPA~\cite{Gleisberg:2003xi} versions
1.0.8 and 1.0.9 and PYTHIA~\cite{Sjostrand:2006za}.  We generated the
SUSY spectrum for the mass points considered using SPheno
2.2.3~\cite{Porod:2003um}. Details regarding the spectrum, cross sections and branching ratios for this point (Point I)
are given in Appendix~\ref{sec:points}.  For the moment, we only need 
to note the resulting masses:
\beq
m_Y=246.6\gev\,,\quad
m_X=128.4\gev\,,\quad m_N=85.3\gev\,.
\label{massesi}
\eeq
We stress however that our
techniques are not specific to SUSY; we have just used the available
tools for supersymmetric models. Thus, event rates employed are not
necessarily those predicted in the context of some particular SUSY
model.  Here, we
simply use a 500 event sample for illustration of the basic ideas.

\begin{figure}[tb]
\begin{center}
 \includegraphics[width=1\textwidth]{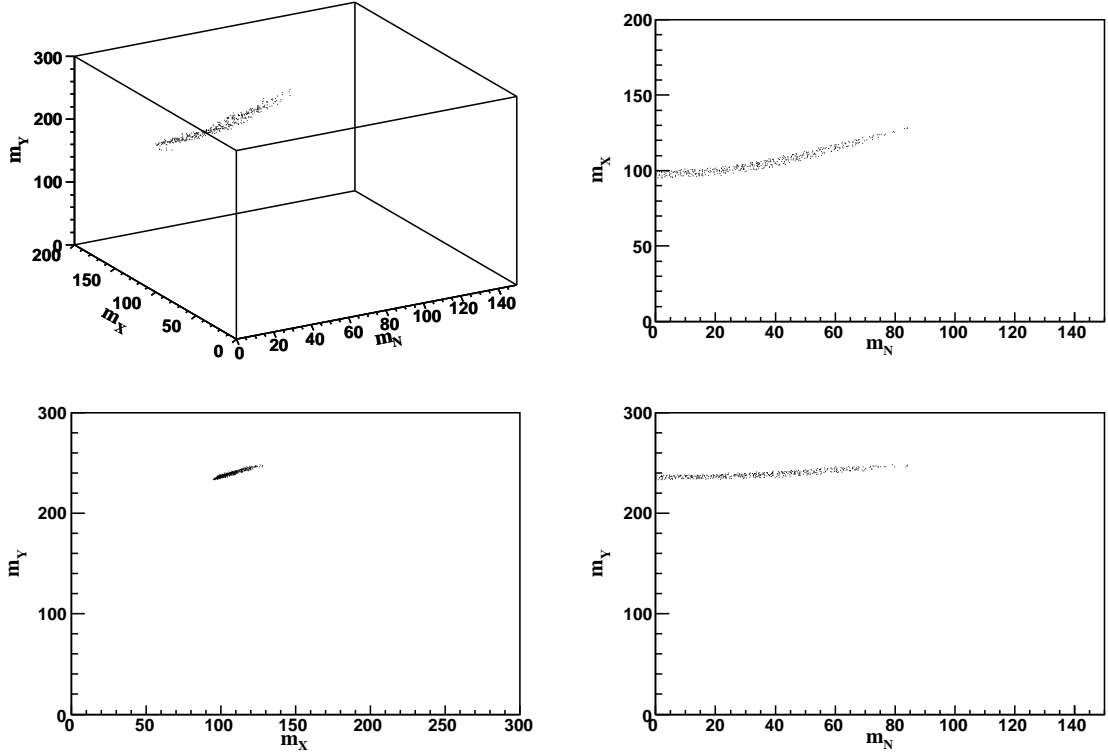}
\end{center}
\caption{\label{fig:nosmear}Mass region (in GeV) that can solve all
  events for the input masses $\{m_Y,m_X,m_N\}=\{246.6,128.4,85.3\}\gev$ using 500 events.  }
\end{figure}

For simplicity, we have required all four leptons to be muons. We assume
that the momenta of the 4 muons and the sum of the transverse momenta of
the two neutralinos are known exactly. The only cuts we have applied on
this sample are acceptance cuts for the muons: $p_T>6 \mbox{ GeV}$ and
$|\eta|<2.5$. We do not consider the  mass of the squark, therefore
information from the quarks is irrelevant except that the presence of
the quark jets typically boosts the system (Fig.~\ref{fig:topology}) away from
the $z$ axis, an effect automatically included in our analysis. In the following, we
denote a set of masses as $\mathcal{M}=\{m_Y,m_X,m_N\}$ and the correct
set as $\mathcal{M}_A$.

Each event defines a mass region in $\mathcal{M}$ space that yields
real solutions for $\vec p_N$ and $\vec p_{N'}$ (for which we often employ
the shorter phrase `real solutions' or simply `solutions').  This region can be
determined by scanning through the mass space.  We then examine the
intersection of the mass regions from multiple events.  This region
must contain the correct masses, $\mathcal{M}_A$.  The allowed mass
region keeps shrinking when more and more events are included. One
might hope to reach a small region near $\mathcal{M}_A$ as long as
enough events are included.  However, this is not the case, as
exemplified in Fig.~\ref{fig:nosmear}. There, the 3-dimensional allowed
region in $\mathcal{M}$-space is shown together with its projections
on 2-dimensional planes. When producing Fig.~\ref{fig:nosmear}, we
discretize the mass space to 1 GeV grids in all three directions. As
already noted, we have used the correct assignments for the locations
of the muons in the decay chains. Wrong assignments will add more
complication; this will be discussed in
Sec.~\ref{sec:with_resolution}. With correct assignments,
and because of our narrow-width  and
no-smearing assumptions, the correct masses $\mathcal{M}_A$ will 
result in at least one real $\vec p_N$ and $\vec p_{N'}$ solution 
for all events and is included in the allowed
region. In all three 2-dimensional projections, 
the entire allowed region is a strip with $m_Y$ and $m_X$
close to the correct values, but $m_N$ left undetermined except for an
upper bound. A lower bound is sometimes present and can be caused by
the presence of events in which the system (Fig.~\ref{fig:topology})
has a large amount of transverse momentum. The upper bound for $m_N$
generally materializes using fewer events than does the lower bound.  By
examining the figures one can see that the upper bound for $m_N$ is
actually close to the correct $m_N$; more generally, $\mathcal{M}_A$
is located near the tip of the cigar shape of acceptable choices in
$\mathcal{M}$-space.  

\begin{figure}[tb]
\begin{center}
\includegraphics[width=0.8\textwidth]{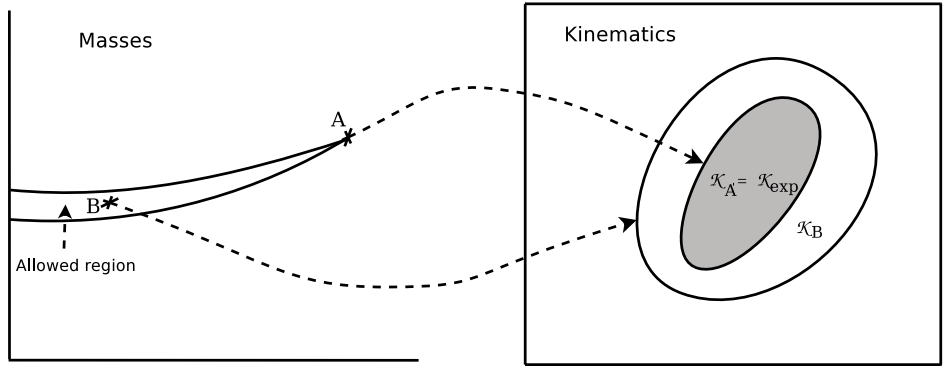}
\end{center}
\caption{\label{fig:map}Map between mass space and kinematic space. The nominal masses, point $A$, produces a kinematic region that coincides with the experimental region:  $\mathcal{K}_A=\mathcal{K}_{exp}$. A point $B$ inside the allowed mass region produces a larger kinematic region: $\mathcal{K}_B\supset \mathcal{K}_{exp}$. }
\end{figure}

An intuitive understanding of why it is that the correct mass set
$\calm_A$ is located at an end point can be garnered from Fig.~\ref{fig:map}.
Any point in the mass space on the left-hand side of the figure is mapped into
a {\it region} of the kinematic space on the right-hand side.  By
`kinematic space' we mean the set of observed 3-momenta of the visible
particles, 3, 4, 5, and 6. 
Thus, the kinematic space has much higher dimensionality than the mass
space --- the on-shell $Y,X,N$ masses can be held fixed while changing the
angles, magnitudes and so forth of the visible particles.  Consequently,
each point in mass space defines a
volume in kinematic space.  In analyzing data, the inverse mapping is
to be envisioned. Each point in the kinematic space
corresponds to a specific momentum configuration of the visible
particles, {\it i.e.} an event. A collection of many events will define a region in the
kinematic space. In particular, the correct set of masses,
point $A$ in Fig.~\ref{fig:map}, produces a kinematic region
$\mathcal{K}_A$ that coincides with the experimental one,
$\mathcal{K}_A=\mathcal{K}_{exp}$, as long as the number of experimental
events is large enough so that all the allowed region is populated.
Any shift away from $A$ will generally not allow one or more kinematical
observables associated with the visible particles to occupy a region
close to the boundary of $\mathcal{K}_{exp}$; {\it i.e.} such a shift will
generally exclude a region of the actually observed
kinematical space.

A mass point other than $\calm_A$ produces a region different from
$\mathcal{K}_{exp}$.  If it does not cover the entire $\mathcal
{K}_{exp}$, this means that some events will not have yielded real
$\vec p_N$ and $\vec p_{N'}$ solutions and, therefore, the mass point
does not appear in the final allowed mass region. On the other hand,
there can be mass points which produce larger kinematic regions
encompassing the entire $\mathcal {K}_{exp}$ region. These mass points
yield real solutions for all events and hence belong to the final
allowed region. This kind of point is exemplified by point $B$ in
Fig.~\ref{fig:map}.  If we shift such a point in the mass space by a small
amount,
$\mathcal{M}_B\rightarrow\mathcal{M}'=\mathcal{M}_B+\delta\mathcal{M}$,
the resulting kinematic region still covers $\mathcal{K}_{exp}$. In
this case, $\mathcal{M}'$ still yields real solutions for all events.
Thus, point $B$, which produces a region larger than
$\mathcal{K}_{exp}$, has the freedom to move in many directions
because it lives inside the allowed region rather than on its
boundary. On the other hand, the correct mass point $A$, which
produces exactly $\mathcal{K}_{exp}$, has the least freedom to move.
In short, locating the correct mass point $\calm_A$ can be viewed as a
kind of generalization of the `edge' method which employs sharp edges
in certain invariant mass combinations constructed from the visible
momenta. Our method matches the whole boundary of the allowed region
in the high-dimensional kinematic space of the visible momenta.

Of course, using the ``tip'' of the allowed mass region is not
applicable in the realistic case where experimental resolutions and
combinatorics are included, not to mention the possible presence of
background events. In particular, some of the events generated after
including these effects will be inconsistent ({\it i.e.} not yield
real solutions for $p_N$ and $p_{N'}$) with the correct mass set
$\mathcal{M_A}$ and so this point will not be contained in the
$\mathcal{M}$ volume obtained if all events are considered.  We must
find more sophisticated methods to identify the correct mass point.
Nevertheless, understanding the idealized case provides useful
guidance for understanding how to deal with the more complicated
realistic situations.

\section{Inclusion of combinatorics, finite resolutions and backgrounds}
\label{sec:with_resolution}

In this section we discuss the more realistic case with finite
resolutions, combinatorics and backgrounds. 
We first discuss the effects from finite
resolutions and combinatorics and later we will include the backgrounds.
For the moment, we continue to employ the spectrum
associated with the SUSY Point I, as specified in
Appendix~\ref{sec:points}, with $\{m_Y,m_X,m_N\}=\{246.6,128.4,85.3\}\gev$.

\subsection{Finite resolution effects and combinatorics}
\label{sub:rescomb}

\begin{figure}[tbh]
\begin{center}
 \includegraphics[width=\textwidth]{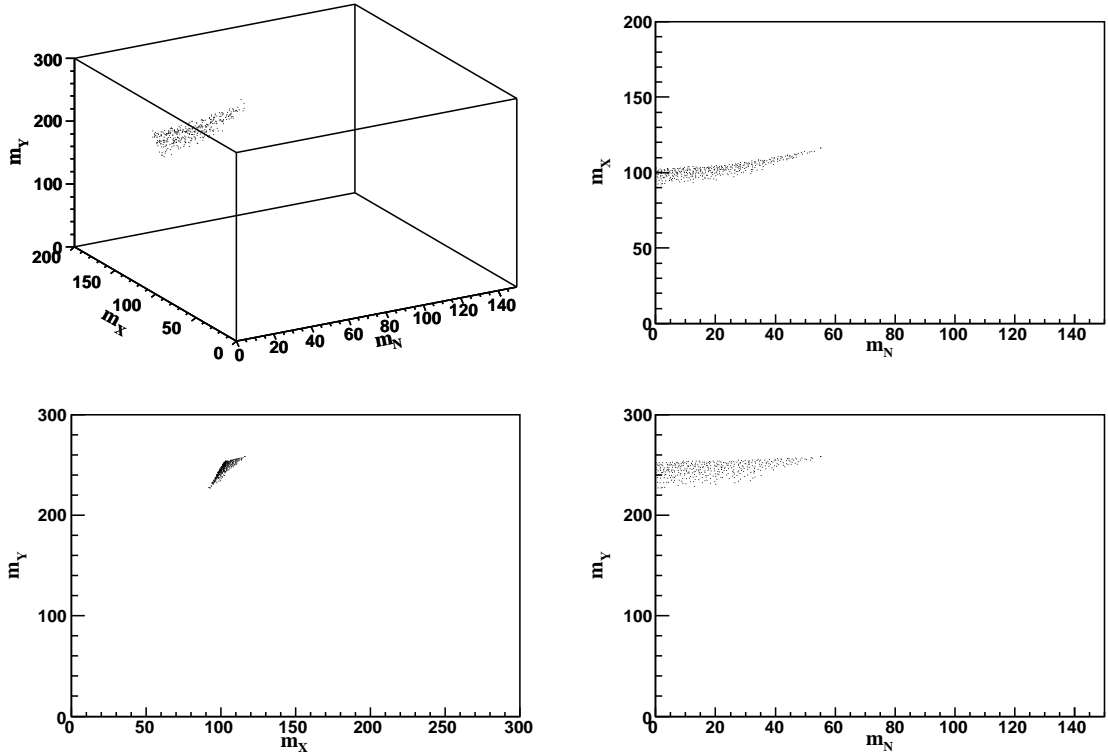}
\end{center}
\caption{\label{fig:smear}The allowed mass region (in GeV) with smearing
and wrong combinatorics. }
\end{figure}

Experimental effects related to smearing and combinatorics will deform
or even kill the allowed mass region. In particular, since the correct
mass point is located at the endpoint, it is most vulnerable to any
mismeasurement. This can be seen in Fig.~\ref{fig:smear}, which
corresponds to 500 events for the same mass point as
Fig.~\ref{fig:nosmear}. The difference is that we have: i) added
smearing; ii) considered all possible combinatoric assignments for the
location of the muons in the two decay chains; and iii) included the
finite widths of the $Y$ and $X$ intermediate resonances.  We smear
muon momenta and missing $p_T$ using the low-luminosity options of the
ATLAS fast simulation package {\tt ATLFAST} as described in Secs. 2.4
and 2.7 of~\cite{atlfast}.  Very roughly, this corresponds to
approximately Gaussian smearing of muon momentum with width $\sim
3\%/p_T$ and of each component of missing momentum $p_T^{\rm miss}$
with width $\sim 5.8\gev$.  We note that we are not directly sensitive
to the resolution associated with individual jet momentum
measurements; uncertainties in the determination of individual jet
momenta are, of course, reflected in the uncertainty of the
determination of $p_T^{\rm miss}$ as we shall shortly review.  Our
approach is only sensitive to $p_T^{\rm miss}$ uncertainties because
we do not look at the jets associated with the chain decays prior to
arriving at the $\cntwo\cntwo$ pair.  We only need the net transverse
momentum of the $\cntwo\cntwo$ pair as a whole, and we determine this
in our procedure as $\sum_{\ell} p_T^{\ell}+\ptmiss$.  Thus, in our
analysis the errors from smearing derive entirely from the
uncertainties in the lepton and missing momentum measurements. The
fact that we don't need to look at individual jets is, we believe, and
important advantage of our approach to determining the $\cntwo$,
$\slep$ and $\cnone$ masses.  Of course, once these masses have been
determined, the edge techniques, which fix mass differences
very accurately, can be used to extract the $\gl$ and $\wtil q$ masses.

We summarize the missing energy procedure as described in Sec.~2.7
of~\cite{atlfast} in a bit more detail. The missing transverse energy
$\etmiss$ is calculated by summing the transverse momenta of
identified isolated photons, electrons and muons, of jets, $b$-jets
and $c$-jets, of clusters not accepted as jets and of non-isolated
muons not added to any jet cluster.  Finally, the transverse energies
deposited in cells not used for cluster reconstruction are also
included in the total sum.  Transverse energies deposited in unused
cells are smeared with the same energy resolution function as for
jets. From the calculation of the total sum $E_T^{\rm obs}$ the
missing transverse energy is obtained, $E_T^{\rm miss}=E_T^{\rm obs}$
as well as the missing transverse momentum components, $p_x^{\rm
  miss}=-p_x^{\rm obs}$ and $p_y^{\rm miss}=-p_y^{\rm obs}$.

For combinatorics, we assume no charge
misidentification. Then, there are 8 independent
possible combinatoric locations for one event,
which can be reduced if one muon pair is replaced by an electron pair.
If any one of these 8 possibilities yields a real solution
(after including smearing/resolution as described above), we include
the $\calm$ point in our accepted mass region.

As regards the resonance widths, these have been computed within the
context of the models we have considered, as detailed in
Appendix~\ref{sec:points}. In our Monte Carlo, the mass of a given
$\cntwo$ or $\slep$ resonance is generated according to a Breit Wigner
form using the computed width.  Although there will be some model
dependence of the widths in that they might differ between the SUSY
models employed as compared to a little-Higgs model, the widths for
these weakly interacting particles are all much smaller than detector
resolutions in both models ({\it e.g.} of order a few hundred MeV in
the SUSY models). This is again an advantage of our approach since we
never need to know where on the Breit-Wigner mass distribution of the
$\gl$ and $\wtil q$ resonances a given event occurs. We only need the
net transverse momentum of the $\cntwo\cntwo$ system as determined
from $\sum_{\ell} p_T^{\ell}+\ptmiss$. Also, for the moment we will
focus on events with four $\mu$'s in the final state and so both
sleptons in the two decay chains will be $\wtil \mu$'s.  When we come
to the SPS1a mSUGRA point, we will discuss combining results for the
$\wtil \mu\wtil\mu$, $\wtil e \wtil e$ and $\wtil \mu \wtil e$ decay
channels. Even in this case, we analyze final states with definite
lepton composition ($4\mu$, $4e$ or $2e2\mu$) separately and do not
need to worry about whether the $\wtil \mu$ is closely degenerate with
the $\wtil e$ (although it in fact is).  If there is significant
non-degeneracy, that would emerge from our analysis. However, to get
final errors on the $\slep$ mass as low as $\sim 5\gev$, degeneracy of
the $\wtil \mu$ and $\wtil e$ must be assumed (and of course is
predicted in the model). If in some other model, the $\wtil \mu$ and
$\wtil e$ are not degenerate, then errors on these individual masses
will be of order $\sim 10-12\gev$, but errors on $\mcnone$ and $\mcntwo$
will  only be slightly larger than the $\sim 5\gev$ level since
the different channel analyzes can be combined for their
determination.

The effects of both wrong combinations and smearing are manifest in
Fig.~\ref{fig:smear}: wrong combinations increase the chance that a
given event can be `solved'~\footnote{We define a  `solved' event to
  be an event
  such that the given $\{m_Y,m_X,m_N\}$ choices yield at least one solution to
  the final quartic equation that leads to physically allowed values
  for $\vec p_N$ and $\vec p_{N'}$. } and therefore broaden the allowed region for
low $m_N$. On the other hand, the allowed region has shrunk in the $m_N$
direction with the new upper bound corresponding to a much smaller
value. This can also be understood by using Fig.~\ref{fig:map}: some
events near the boundary of $\mathcal{K}_A$ can be resolution-smeared 
to a location outside of
$\mathcal{K}_A$,  which renders $\mathcal{K}_{exp}$ larger than
$\mathcal{K}_A$. Thus the correct mass point $A$ is removed from the
allowed mass region. For point $B$ which corresponds to a larger
kinematic region, if the  fluctuation is small enough,
$\mathcal{K}_{exp}$ is still within $\mathcal{K}_B$ and therefore does
not disappear.  Of course, if the smearing is large, the entire allowed
region can be eliminated. The effect from background events, as
considered in the next subsection, will be 
similar. Since background events are produced
by some completely different processes there is no reason to expect
that multiple background events can be solved by the assumed topology
with a given choice of $\calm$. Thus, background events tend to reduce the allowed
region.

\begin{figure}

\begin{center}
 \includegraphics{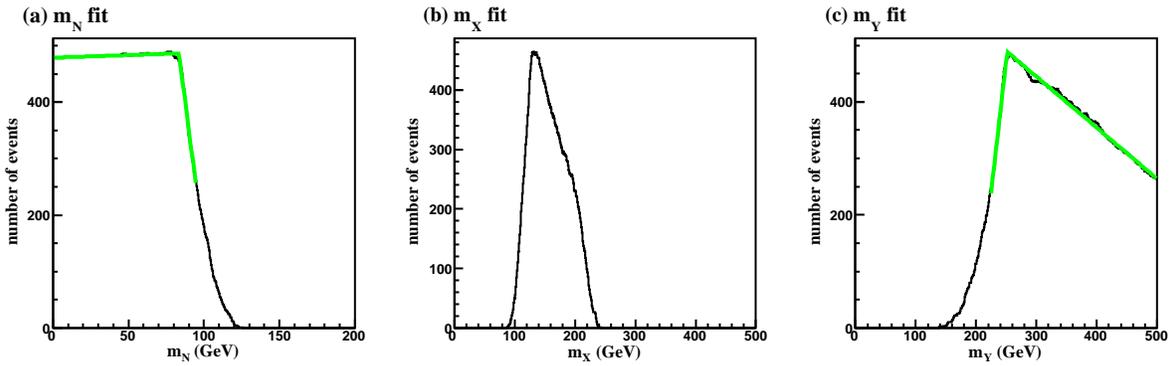}
\caption{\label{fig:fits}One-dimensional fits by fixing the other two
masses at the correct values. }
\end{center}
\end{figure}

From the above observation, one concludes that allowed mass region in
general does not exist and, even if it exists, we can not read directly
from it the correct masses. Some other strategy must be employed. An
obvious choice is to examine the number of solvable events for various
given masses.  We can not simply maximize the number of solvable events
and take the corresponding masses as our estimate---such a procedure would still favor
low $m_N$ values.  Instead, we choose to look for the mass location where the
number of solvable events changes drastically. This kind of location is
most easily illustrated in one dimension. For example, in
Fig.~\ref{fig:fits} a, we fix $m_Y$ and $m_X$ to the correct (input)
values, and count the number of solvable events as a function of $m_N$.
(In this figure and the following discussion, we use bin size of 0.1
GeV). A sudden drop around the correct $m_N$ is obvious. Similarly,  in
Figs.~\ref{fig:fits} b and  \ref{fig:fits} c we have fixed $m_Y$ and $m_N$
($m_X$ and $m_N$) and also see clear ``turning points'' near the
correct $m_X$ ($m_Y$) mass. 
To pin down where the turning points are located, we fit
Figs.~\ref{fig:fits} a and \ref{fig:fits} c to two
straight line segments and take the intersection point as the turning
point. 

We can not fix {\it a priori} two of the masses to the correct values
since they are unknown. On the other hand, to search for the sharpest
turning point directly in the 3-dimensional space is numerically
non-trivial. This observation motivates us to obtain the masses from a
series of one-dimensional fits.  We start from some random set of
masses and carry out a recursive series of one-dimensional fits to the
number of solved events as a function of $m_N$, $m_X$ or $m_Y$ holding
$\{m_Y,m_X\}$, $\{m_Y,m_N\}$, or $\{m_X,m_N\}$ fixed, respectively.
Each such one dimensional fit gives us a sharp turning point that is
used to set an updated value for $m_N$, $m_X$ or $m_Y$, respectively.
We use this new value in performing a fit for the next mass in the
sequence in the next step. One might hope that this procedure will
converge to the correct mass values, but in practice, even though the
procedure passes through the correct mass point, the fitted masses
keep increasing and the recursion does not stabilize at the correct
mass point.  However, as we will see, there is a simple way to get the
masses out of the fits.

Having discussed the main ingredients of the method, we present a
specific procedure for obtaining the masses.  The procedure is applied
to a data sample corresponding to $90\fbi$ at the LHC, using
the event rates and branching ratios obtained for the SUSY Point I as
detailed in Appendix~\ref{sec:points}, which, in particular, gives 
the same masses as those employed in Sec.~\ref{sec:no_resolution}:
$\{m_Y,m_X,m_N\}=\{246.6,128.4,85.3\}\gev$. 
Taking into consideration the decay
branching ratios, the number of events is
roughly 2900. In order to mimic reality as much as possible, 
experimental resolutions and wrong
combinatorics are included.  To reduce the SM background, we
require that all muons are isolated and pass the kinematic cuts:
\begin{equation}
   \label{cuts}
   |\eta|_\mu<2.5, \;\; p_{T\mu}>10~\mbox{GeV},\;\; {\slash\hspace{-0.25cm} p}_T>50~\mbox{GeV}.
\end{equation}
With these cuts, the four-muon SM background is negligible.  The
number of signal events is reduced from 2900 to about 1900. 

\begin{figure}[tbh]
\begin{center}
 \includegraphics  {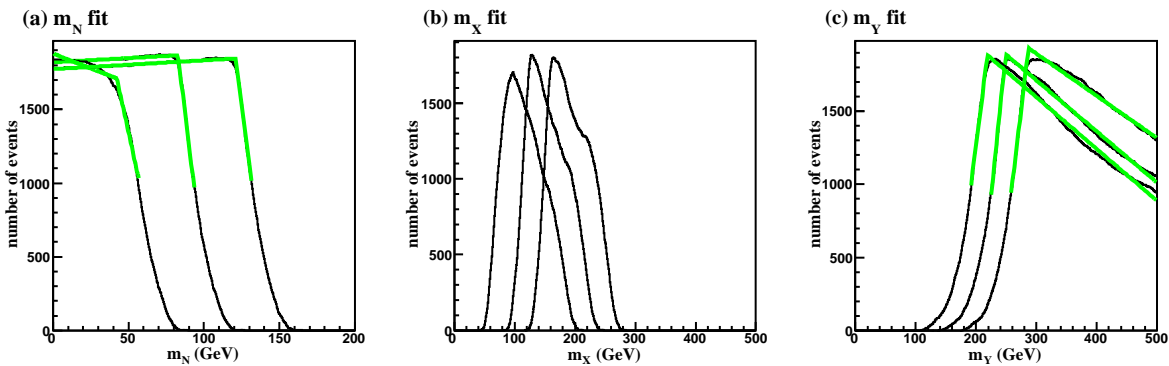}
\caption{\label{fig:steps}A few steps showing the migration of the one
dimensional fits. The middle curve in each plot corresponds to masses
close to the correct values.}
\end{center}
\end{figure}

\begin{figure}[tbh]

\begin{center}
 \includegraphics[width=0.6\textwidth]{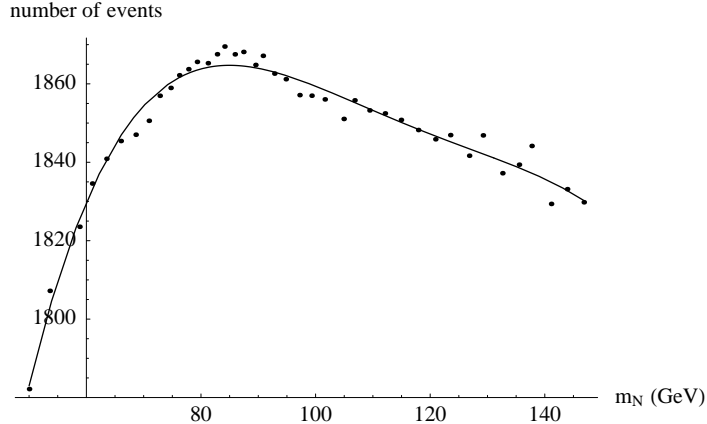}
\caption{\label{fig:mndetermine}The final plot for determining $m_N$.
The position of the maximum of the fitted polynomial is taken to 
be the estimation of $m_N$.}
\end{center}
\end{figure}

The procedure comprises the following steps:
\begin{enumerate}
\item \label{step1}Randomly select masses $m_Y>m_X>m_N$ that are below the correct masses (for example, the current experimental limits).
\item \label{step2}Plot the number of solved events, $N_{evt}$, as a function of
  one of the 3 masses in the recursive order 
  $m_N$, $m_X$, $m_Y$  with the other two masses fixed. In the case of
  $m_Y$ and $m_N$, we fit $N_{evt}$
  for the plot with two straight lines and adopt the mass value at
  the intersection point as the updated mass.  In the case of
  $m_X$, the updated mass is taken to be the mass at the peak of
  the $N_{evt}$ plot.

A few intermediate one-dimensional fits are shown in Fig.~\ref{fig:steps}.

\item \label{step3}Each time after a  fit to $m_N$, record the
  number of events at the intersection (sometimes called the turning
  point) of the two straight lines, as exemplified 
  in Fig.~\ref{fig:fits} a. This event number at the turning point
  will in general be non-integer.
 
\item \label{step4}Repeat steps \ref{step2} and \ref{step3}. The
  number of events recorded  in step \ref{step3} will in general
  increase at the beginning and then decrease after some steps, as
  seen in Fig.~\ref{fig:mndetermine}. Halt the recursive procedure
  when the number of (fitted) events  has sufficiently passed the maximum position.  

\item \label{step5}Fit Fig.~\ref{fig:mndetermine} to a (quartic) polynomial and take the position where the polynomial is maximum as the estimated $m_N$.

\item \label{step6}Keep $m_N$ fixed at the value in step \ref{step5} and do a few one-dimensional fits for $m_Y$ and $m_X$ until they are stabilized. Take the final values as the estimates for  $m_Y$ and $m_X$.
\end{enumerate}

\begin{figure}
\begin{center}
 \includegraphics[width=0.8\textwidth]{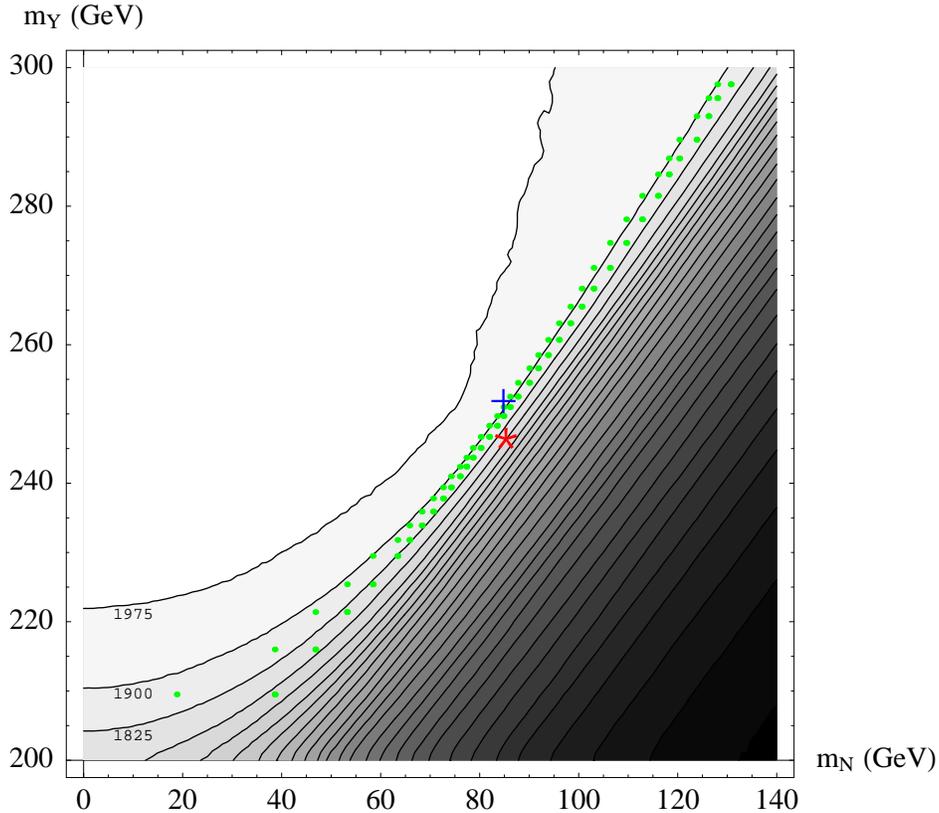}
\caption{\label{fig:contour}Contours for the number of solved events in
the $m_N\sim m_Y$ plane with 2000 events. The number of events is the
maximum value obtained after varying $m_X$. Contours are plotted at
intervals of 75 events, beginning with a maximum value of 1975.
The red star is the position for the correct
masses and the blue cross is the position of the fitted masses. The
green dots correspond to a set of one-dimensional fits. }
\end{center}
\end{figure}

A deeper understanding of our procedure can be gained by examining the
graphical representation of the steps taken in the $(m_Y,m_N)$ plane
shown in Fig.~\ref{fig:contour}. There, we display contours of the
number of (fitted) events after maximizing over possible $m_X$
choices. The contours are plotted at intervals of 75 events, beginning
with a maximum value of 1975 events.  As we go from 1975 to 1900 and
then to 1825 events, we see that the separation between the contours
decreases sharply and that there is a `cliff' of falloff in the number
of solved events beyond about 1825 events. It is the location where
this cliff is steepest that is close to the input masses, which are
indicated by the (red) star.  The mass obtained by our recursive
fitting procedure is indicated by the (blue) cross.  It is quite close
to the actual steepest descent location. It is possible that use of
the contour plot by visually picking the point of steepest descent
might also yield an accurate mass determination comparable to or
possibly even superior to that obtained (and specified in detail
below) using the recursive fitting procedure.  Roughly, the
steepest descent point corresponds to the point where the magnitude of
$\vec\nabla^2$ in mass space is maximized.  Unfortunately, even after
some smoothing, the second derivative is quite `noisy' and therefore
not particularly useful in a local sense.  The one-dimensional fits
give us a quick and intuitive way to find this maximum, and the
associated recursive procedure has the advantage of being insensitive
to statistical fluctuations in the number of events at a single point.
Of course, if one has the computer power,
probably the most accurate procedure would be to directly fit the
3-d $N_{evt}$ vs. $\{m_Y,m_X,m_N\}$ histogram.  Fig.~\ref{fig:contour}
is constructed from a 1-d projection of the 3-d space, and has
therefore lost some information.

Following the recursive fitting procedure, the final values for the
masses are determined to be \{252.2, 130.4, 85.0\} GeV, which are all
within a few GeV of the actual input values, \{246.6, 128.4, 85.3\}
GeV.  The procedure is empirical in the sense that many of the steps
could be modified and improved.  In particular, above we adopted the criterion
that the correct masses maximize the number of events at the turning
points in the $m_N$ fits, which is justified by Fig.~\ref{fig:steps}
a.  Instead, we might opt to maximize the number of events in the
$m_X$ fits shown in Fig.~\ref{fig:steps} b. One could also change the
order of fits in step \ref{step2} and change the fit function from
straight lines to more complicated functions, {\it etc.}  We have
tried several different strategies and they yield similar results.
Finally, one could simulate the signal for a mass point and directly
generate Fig.~\ref{fig:steps}, changing the masses until we get the
best possible fit to the data; but, this is very computationally
intensive.

The recursive procedure does not provide an easy way to evaluate the
errors in the mass determination.  For this purpose, we generate 10
different data samples and apply the procedure to each sample. As above,
each sample corresponds to 1900 
experimental data points after cuts. Then, we
estimate the errors of our method by examining the statistical
variations of the 10 samples. This yields central masses and rms
errors of
\begin{equation}
m_Y=252.2\pm 4.3 ~\mbox{GeV},\ \ m_X=130.4\pm 4.3~\mbox{GeV},\ \ m_N=86.2\pm 4.3~\mbox{GeV}.
\end{equation}
The statistical variations for the mass differences are much smaller:
\begin{equation}
m_Y-m_X=119.8\pm 1.0 ~\mbox{GeV},\ \ m_X-m_N=46.4\pm 0.7 ~\mbox{GeV}.
\end{equation}
Compared with the correct values, $\mathcal{M}_A=\{246.6, 128.4,
85.3\}\gev$, we observe small biases in the mass determination,
especially for the mass differences, which means that our method has
some ``systematic errors''.  (The biases will, of course, depend upon
the particular functions employed for the one dimensional fits --- our
choice of using straight lines is just the simplest.)
One technique for determining the biases
is to perform our analysis using Monte Carlo data. In particular, one
could examine the plots of number of `solved' events vs. test mass as
obtained from the data vs. those obtained from a Monte Carlo in which
definite input masses (which are distinct from the test masses
employed during our recursive procedure) are kept fixed. One would
then search for those input masses for which the distributions of the
solved event numbers from the Monte Carlo match those from the data.
Knowing the underlying Monte Carlo masses as compared to the extracted
masses would allow us to subtract the differences, thereby removing
the biases. This procedure would not appreciably change the errors
quoted above. We believe that the biases are mainly a function of the
underlying masses and broad kinematic event features.  However, there
may be some weak dependence of the biases on the actual model being
employed.  Within the context of a given, {\it e.g.} SUSY, model, the
bias can be quite accurately determined.

In the above error estimation, we have neglected the uncertainties
coming from varying the choice of the starting point in mass space
used to initiate the recursive sequence of fits.
This may introduce an error for the absolute mass
scale of order the step size around the correct masses. For the masses
chosen, it is about 1 GeV and much smaller than the uncertainties from
varying data samples.

The reader may be surprised at the small size of the errors quoted
above given that the error in the measurement of the missing momentum
of any one event is typically of order $5\gev$ or larger. The
explanation is similar to that associated with understanding the small
errors for the edge locations in the edge approach. 
In the edge approach,
the location of the edge for some mass variable $m_{\rm vis}$ is
obtained by fitting data obtained at several $m_{\rm vis}$ values.
Each such data point has many contributing events and the average
value will obviously have much smaller error than the value for any
one contributing event. The fit to the edge will further reduce
sensitivity to individual events.  In our approach, the edge in the
distribution of $N_{evt}$ as a function of one of the trial masses
($m_N$, $m_X$ or $m_Y$) will similarly be an average over many events
and the uncertainty of the location of this edge will be much smaller
than the uncertainties in the measurements of the lepton momenta and
missing momentum of any one event.

\subsection{Backgrounds}
\label{sub:backgrounds}

For the point we have chosen with a 4 muon + missing energy final state,
the background is negligible.  We examined backgrounds arising from
$ZZZ$, $ZWW$, $t\bar{t}$, $t\bar{t}Z$, $t\bar{t} b\bar{b}$, and
$b\bar{b}b\bar{b}$.  Muons from bottom and charm decays are never very
hard nor  isolated, and can be easily separated from the signal with
basic isolation criteria.  Tri-boson production simply has tiny cross
sections, especially after requiring all-leptonic decays.

\begin{figure}
\begin{center}
 \includegraphics{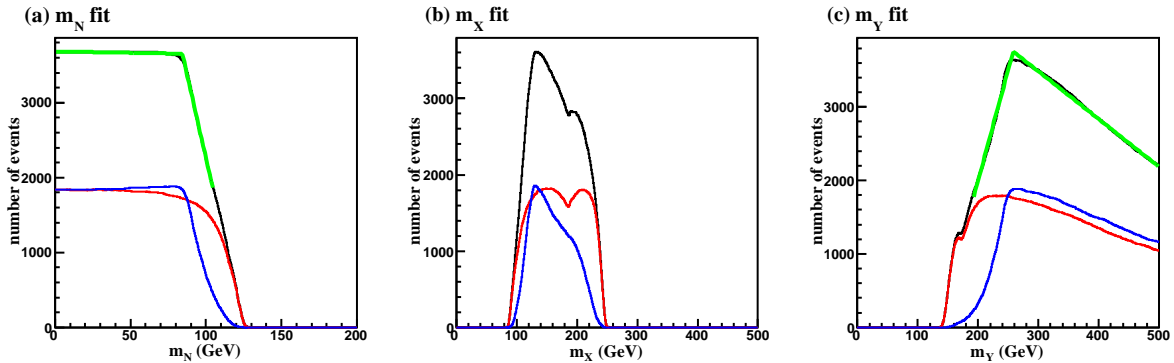}
\caption{\label{fig:withbgs}Fits with 1900 
  signal events (after cuts) and an equal
  number of background events. Separate numbers of signal (blue) and
  background (red) events are also shown. } 
\end{center}
\end{figure}

Thus, we must `artificially' introduce background in order to see what
its effect might be on our procedures.  For this purpose, we 
generate $t\bar{t}$ events, where the
$W$'s decay to muons.  We require that the $b$ quarks decay to muons,
{\it but do not require them to be isolated.}  In many ways,  this is a
near-worst case background since it has a similar topology 
aside from the final $b\to \mu+\ldots$ decays. However, the missing
neutrinos imply that the missing momentum may be significantly
different.  As noted, this is not a realistic
background as it could be removed by simple isolation cuts on the muons.

\begin{figure}
\begin{center}
 \includegraphics[width=0.6\textwidth]{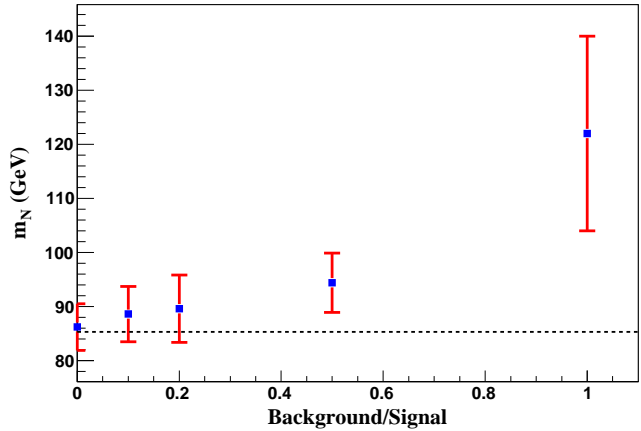}
\caption{\label{fig:errors} $m_N$ determination with different
background-signal ratio. The dashed horizontal line corresponds to the
correct $m_N$.}
\end{center}
\end{figure}

Adding a number of background events equal to the number of signal
events, {\it i.e.} 1900 events after cuts, we repeat the
one-dimensional fits. A typical cycle around the correct masses is
shown in Fig.~\ref{fig:withbgs}. For comparison the numbers of
solvable signal events and background events are also shown
separately.  The effect of background events is clear: the curve for
solvable background events is much smoother around the turning point,
and therefore smears but does not destroy the turning point.  Although
we are considering one specific background process, this effect should
be generic, unless the backgrounds happen to have non-trivial features
around the turning points. Nevertheless, due to the fact that there
are 8 possible combinatoric muon locations, the chance that a
background event gets solutions is quite large and they do affect the
errors and biases of the mass determination.  This can be seen in
Fig.~\ref{fig:errors}, in which we have used the same 10 sets of
signal events as in the previous subsection, but varied the number of
background events according to the ratio $B(ackground)/S(ignal) =0,
0.1, 0.2, 0.5, 1$.  We observe increases in both the biases and
variations about the central values.  For $B/S\ge 1$, the maximum in
the $m_N$ determination is obscured or even lost and we start to get
random results. For $B/S\lsim 0.2$, we are close to the $B=0$ results.

It is important to emphasize that the above analysis is pessimistic in that it
assumes that we do not understand the nature/source of
the background events. One procedure that could significantly improve
on uncertainties associated with the background would be to 
Monte Carlo the background or use
extrapolations of measured backgrounds ({\it e.g.} those as measured
before the cuts that make the signal a dominant as compared to a small
component of the observed events)
and then apply our recursive procedure to the
known background and at each stage subtract off the background
contribution to a given plot of the number of events vs. $m_N$, $m_Y$
or $m_X$. After such subtraction, the recursive procedure will yield
essentially identical results to that obtained  in the absence of
background unless the background itself is not smooth in the vicinity
of the `turning' points.  

The importance of finding cuts that both select a given topology and
minimize background is clear. If it should happen that the we assume
the wrong topology for the events retained, then our analysis itself
is likely to make this clear.  Indeed, events with the ``wrong''
topology would almost certainly yield a smooth distribution in plots of
retained event number vs. any one of the masses of the resonances
envisioned as part of the wrong topology. It is only when the 
correct topology is employed that sharp steps will be apparent in all
the event number vs. resonance mass plots.

Another important situation to consider is that in which it is
impossible to find a set of cuts that isolates just one type of decay
topology, so that there are several signal processes contributing
after a given set of cuts. 
However, it is quite easy to find situations where there are different
signal processes yielding very similar final decay topologies, all of
which would be passing through our analysis.  One must then look for
additional tricks in order to isolate the events of interest. In some
cases, this is possible on a statistical, but not event-by-event
basis. The SPS1a SUSY point provides an interesting example that we
will consider shortly.

\section{Other processes and mass points}
\label{sec:otherpoints}

Our method is generic for the topology in Fig.~\ref{fig:topology}, 
and in particular is not
restricted to the SUSY process we have considered so far.  The
statistical variations and biases probably do depend 
to some extent on the process.
For example, if the visible particles 5 (6) and 3 (4)  are of different
species, the number of wrong combinatorics will be  reduced and we
would expect a better determination of the masses. On the other hand, if one
or more of the visible particles are jets, the experimental resolution
and therefore the statistical error will be worse than in the 4-lepton
case.  
\subsection{Changing relative mass differences in the chain}
\label{sub:small}

\begin{figure}
\begin{center}
 \includegraphics{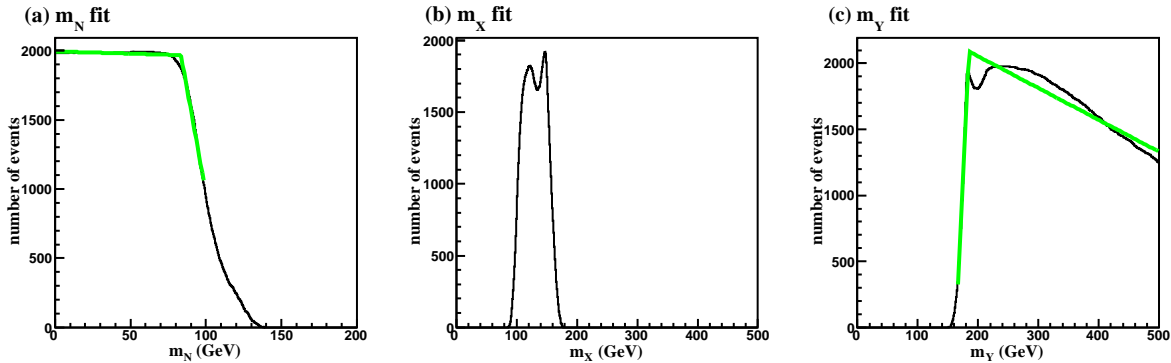}
\caption{\label{fig:point2}One-dimensional fits for mass point \{180.8, 147.1, 85.2\} GeV. }
\end{center}
\end{figure}

The errors in the mass determination also depend on the mass point,
especially the two mass differences, $\Delta m_{YX}=m_Y-m_X$ and
$\Delta m_{XN}=m_X-m_N$. In Fig.~\ref{fig:point2}, a set of
one-dimensional fits are shown for mass point $\calm=\{180.8, 147.1, 85.2\}$
GeV (which we label as Point II). We will assume 2000 events after
cuts, very similar to the 1900 events 
remaining after cuts for Point I.  Point II differs from Point I 
in that for Point II 
$\Delta m_{YX}<\Delta m_{XN}$, while for Point I $\Delta m_{YX}>\Delta
m_{XN}$. The double peak structure in the Point II $m_X$ fit
(Fig.~\ref{fig:point2} b) is evident. The curve to the right of the
turning point in Fig.~\ref{fig:point2} c is also ``bumpy'' compared
with Fig.~\ref{fig:fits} c. These features are induced by wrong
combinatorics. In the process we consider, all visible particles are
muons, so they could be misidentified as one another and still yield
solutions.  Roughly speaking, $\Delta m_{YX}$ and $\Delta m_{XN}$
determine the momentum of the particles 5 (6) and 3 (4) in
Fig.~\ref{fig:topology}, respectively.  Therefore, the chance that a
wrong combinatoric yields solutions is enhanced when, for example,
$\Delta m_{YX}$ is close to the correct value of $\Delta m_{XN}$.
When the two mass differences are close to each other, the turning
point is smeared.  Nonetheless, with 2000 events after cuts,
the errors obtained for the masses are similar to those obtained for
Point I.

\subsection{Small LSP mass}
\label{sub:smalllsp}

\begin{figure}
\begin{center}
 \includegraphics[width=0.6\textwidth]{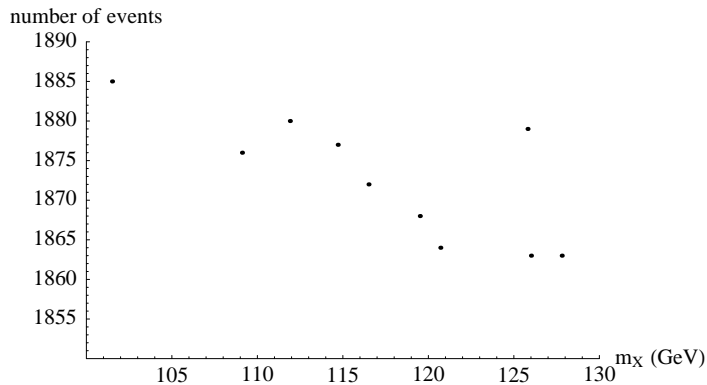}
\caption{\label{fig:mnu0}Peak number of events in $m_X$ fits for  mass
point \{199.4, 100.0, 0.1\} GeV. }
\end{center}
\end{figure}

Another interesting case is that of $m_N$ being zero or very small. 
As for the previous case, we have arbitrarily used a sample of 2000
events after cuts. Because
the one-dimensional fits proceed in the direction of increasing masses,
we will miss the correct masses even when we start from $m_N=0$.  Since
we always fit the $m_N$ plot to two line segments, it will never yield
$m_N=0$. However, we can distinguish this case by looking at the peak
number of events in the $m_X$ fits. For example, considering mass point
\{199.4, 100.0, 0.1\} GeV (which we call Point III),  we start from $m_X=80.0$ and $m_N=0.0$ and
fit the masses in the order: $m_Y\rightarrow m_X\rightarrow m_N$.  The
first few fits yield $$\{205.0 ,   80.0, 0\}\rightarrow\{205.0 , 101.5,
0\}\rightarrow\{ 205.0, 101.5, 24.6\}\rightarrow\cdots$$ After only two
steps, the $Y$ and $X$ masses are adjusted close to the correct
values.  Examining 
the peak number of events in the $m_X$ fits (Fig.~\ref{fig:mnu0}), we
find that the number is maximized in the first $m_X$ fit. This is
clearly different from previous cases where the number of events always
increases for the first few $m_X$ fits (see Fig.~\ref{fig:steps} b), and
indicates that $m_N$ is near zero.  

\subsection{The SPS1a Point}
\label{sub:sps1a}

\begin{figure}
\begin{center}
 \includegraphics[width=0.6\textwidth]{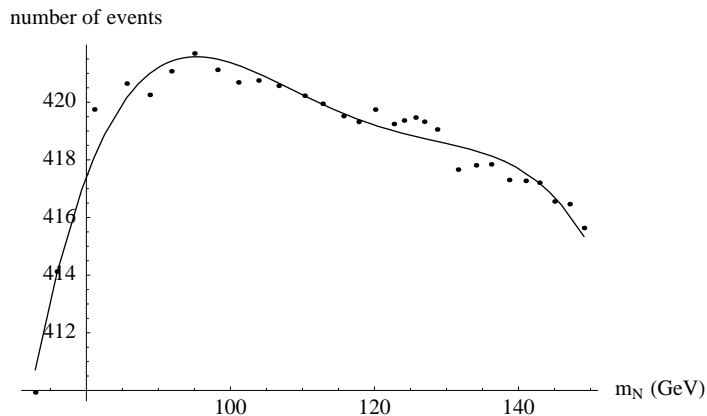}
\caption{\label{fig:sps1a} Fitted number of events at the turning
  point as a function of $m_N$ for the fits for  the SPS1a case.}
\end{center}
\end{figure}

It is desirable to compare directly to the results obtained by
others for the SPS1a SUSY parameter point.  We perform the analysis
using the same $4\mu \cnone\cnone$ final state that we have been
considering.  For the usual SPS1a mSUGRA inputs (see
Appendix~\ref{sec:points}) the masses for $Y=\cntwo$, $X=\wt \mu_R$ and
$N=\cnone$ (from ISAJET 7.75) are $180.3\gev$, $142.5\gev$ and
$97.4\gev$, respectively. This is a more difficult case than Point I considered earlier
due to the fact that the dominant decay of the $\cntwo$
is $\cntwo\to \tau\stauone$. The branching ratio for $\cntwo\to \mu
\wt \mu_R$ is such as to leave only about $1200$ events in the 
$4\mu\cnone\cnone$ final state after $L=300\fbi$
of accumulated luminosity. Cuts reduce the number of events further to
about 425. This is too few for our technique to be as successful as
for the earlier considered cases. 
After including combinatorics and resolution we obtain:
\beq
m_Y=188\pm12\gev\,, \quad m_X=151\pm 14\gev\,,\quad m_N=100\pm 13\gev\,.
\eeq
In Fig.~\ref{fig:sps1a}, we give an SPS1a plot analogous to
Fig.~\ref{fig:mndetermine}.  Errors are determined by generating many
such plots for different samples of 425 events. Note the vertical scale.
The change in the number of events as one varies $m_N$ is quite small
for small event samples and this is what leads to the larger errors in
this case.

In principle, we must also take into account the fact that the $\cntwo \to
\tau\stauone$ decays provide a background to the purely muonic final
state.  The dominant decay $\cntwo\to \tau \stauone$ has a branching
ratio that is a factor of $\sim 14$ times larger than that for
$\cntwo\to\ell\wt \ell_R$.~\footnote{This is, of course, due to the
  fact that $\cntwo$ prefers to couple to left-handed slepton
  components, which are significant for the $\stauone$.}  The
$\stauone$ will then decay to $\tau\cnone$.  If both $\tau$'s then
decay to $\mu \nu\anti \nu$, then $\cntwo\to \tau\stauone$ events will
be likely to contaminate the $\cntwo \to \mu\wt\mu_R$ sample.
Fortunately, this contamination is not huge.  The relevant effective
branching ratios 
for $\cntwo\cntwo\to \tau\stauone\tau\stauone\to 4\mu\cnone\cnone$ and
$\cntwo\cntwo\to \tau\stauone\mu\wt\mu_R\to 4\mu\cnone\cnone$ are
\bea
&&\left[{\br(\cntwo\to \tau\stauone \to\tau\tau\cnone\to \mu\mu
    4\nu\cnone)\over \br(\cntwo\to \mu \wt \mu_R\to \mu\mu
    \cnone)}\right]^2
\sim \left[14 \times (0.174)^2\right]^2\sim 0.18
\eea
and
\bea
&&2\left[{\br(\cntwo\to \tau\stauone \to\tau\tau\cnone\to \mu\mu
    4\nu\cnone)\over \br(\cntwo\to \mu \wt \mu_R\to \mu\mu
    \cnone)}\right]\sim 0.85\,, 
\eea 
respectively.  The contamination levels from these backgrounds are further reduced by factors
of $\sim 5$ for the $\cntwo\cntwo\to \tau\stauone\tau\stauone$ final
state and by $\sim 2$ for the $\cntwo\cntwo\to
\tau\stauone\mu\wt\mu_R$ final state after imposing the simple cuts of
Eq.~(\ref{cuts}) (due to the softer nature of the $\mu$'s coming from
the $\tau$ decays), implying contamination at about the $3.6\%$ and
$40\%$ levels, respectively.  Clearly, it is important to reduce this
level of contamination given that $\mstauone$ is smaller than
$\mslepr$ by about $15\gev$ and so, to the extent that events
containing $\cntwo\to \tau \stauone$ decays remain in our sample, they
might contribute additional structures to our plots of number of
events vs.  mass.  This reduction can be accomplished on a statistical basis
using a further trick analogous to that discussed (but not, we believe,
actually employed) in Ref.~\cite{Gjelsten:2004ki}.  They note that the
decay sequences $\cntwo\to \mu^- e^+\cnone$ and $\cntwo\to \mu^+
e^-\cnone$ are unique to $\cntwo\to \tau\stauone$.  Thus, when
considering just the one-sided decay chain situation one can subtract
off (on a statistical basis, {\it i.e.} after many events) the
$\cntwo\to \tau \stauone$ background by 
\beq N(\cntwo\to \mu\wt
\mu_R\to \mu \mu \cnone)=N(\cntwo\to \mu\mu \cnone)-N(\cntwo\to \mu
e\cnone)\,,
\eeq 
where $N$ is the number of `solved' events as a function of one of the
unknown on-shell masses. In our case, where both chain decays are considered
simultaneously, we have $4\mu\cnone\cnone$ states arising from
$\cntwo\cntwo\to \tau^\pm\stauone^\mp\tau^\pm\stauone^\mp$ decays and
$\cntwo\cntwo\to \tau^\pm\stauone^\mp \mu^\pm \wt\mu_R^\mp$ decays
in addition those from our $\cntwo\cntwo\to \mu^\pm \wt \mu_R^\mp\mu^\pm\wt \mu_R^\mp$
signal.
To subtract off the background SUSY events from the former two decay chains, we can employ the
following subtraction (where the initial $\cntwo\cntwo$ and final
$\cnone\cnone$ are implicit)
\bea N(\mu^\pm\wt \mu_R^\mp \mu^\pm\wt
\mu_R^\mp\to \mu^+\mu^-\mu^+\mu^-) &=& N(\mu^+\mu^-\mu^+\mu^- )
- N(e^+\mu^-\mu^+\mu^-)+N(e^+e^+\mu^-\mu^-)\nonumber\\
&=& N(\mu^+\mu^-\mu^+\mu^- )
-\quarter\Bigl[N(e^+\mu^-\mu^+\mu^-)+N(e^-\mu^+\mu^-\mu^+)\nonumber\\
&&\qquad\qquad\qquad\qquad +N(\mu^+e^-e^+e^-)+N(\mu^-e^+e^-e^+)\Bigr]\nonumber\\
&&\qquad
\qquad\qquad\quad+\half\left[N(e^+e^+\mu^-\mu^-)+N(e^-e^-\mu^+\mu^+)\right]\,.
\nonumber\\
\eea 
where the latter form is likely to have smaller statistical error.  An
experimental indicator of the sensitivity to statistics could be
gained by examining the different possible equivalent subtractions, of
which only two are indicated above.  If one were happy to ignore the
$3.6\%$ contamination from $\cntwo\cntwo\to
\tau^\pm\stauone^\mp\tau^\pm\stauone^\mp$ decays one could then use a
simpler form to subtract off the dominant contamination from
$\cntwo\cntwo\to \tau^\pm\stauone^\mp \mu^\pm \wt\mu_R^\mp$ decays,
namely
\bea N(\mu^\pm\wt \mu_R^\mp \mu^\pm\wt \mu_R^\mp\to
\mu^+\mu^-\mu^+\mu^-) &\sim &
N(\mu^+\mu^-\mu^+\mu^- ) - N(e^+\mu^-\mu^+\mu^-)\nonumber\\
&\sim & N(\mu^+\mu^-\mu^+\mu^- )
-\half\left[N(e^+\mu^-\mu^+\mu^-)+N(e^-\mu^+\mu^-\mu^+)\right]\,.
\nonumber\\
\eea
We have not actually performed this kind of analysis using
any of the possible subtractions to see how
well we do, but we expect that the net background contamination will
be equivalent to $B/S\lsim 0.1$, a level for which our techniques work
very well and the errors quoted earlier for the SPS1a point using the
$4\mu$ final state will not
be increased by very much. 

Of course, the same analysis as performed for the $4\mu$ final state
can also be used for the $2\mu 2e$ and $4e$ final states.
Combinatorics are less of an issue for the $2\mu 2e$ final state than
for the $4\mu$ and $4e$ final states.  The $4e$-channel event number
is essentially the same as the $4\mu$-channel event number and the
$2\mu 2e$-channel event number is roughly twice as large.  Combining
all channels (as appropriate if the $\wtil e$ has mass very close to
the $\wtil \mu$, as predicted by the model), one obtains a total of
about 1700 events and $\sim 5\gev$ errors for our $\mcntwo$, $\mslep$
and $\mcnone$ determinations.

Of course, as the observant reader may have noticed, to get 1700
events requires running at high luminosity, whereas the simulations
referenced so far have employed the $\ptmiss$ resolution expected at
low-luminosity running.  The $\ptmiss$ low-luminosity resolution is
about $5.8\gev$ and that at high luminosity is about $11.6\gev$.
However, we have argued that perhaps one is not all that sensitive to
this resolution when considering a large collection of events and
looking for the location of an edge in the number of reconstructed
events. We have used the SPS1a point to test this hypothesis by
repeating our analysis using high-luminosity running resolutions.

The results confirm our hypothesis. First, the worse resolution
results in our accepting somewhat more events than previously, roughly
480 events (this is the average for the  10 Monte Carlo
``experiments'' employed) in the $4\mu$ channel for the same cuts (and $L=300\fbi$).
The resulting mass determinations obtained using the 10 independent
Monte Carlo experiments are
\beq
m_Y=187\pm10\gev\,, \quad m_X=151\pm 10\gev\,,\quad m_N=98\pm 9\gev\,,
\eeq
where the errors are, as always, rms errors.
In short, we get even smaller errors than for low-luminosity running.
After combining the $4\mu$, $4e$ and $2\mu2e$ channels assuming $\wtil
e$--$\wtil \mu$ degeneracy our mass determination errors are slightly
above $4\gev$.

\section{Summary and Discussion}
\label{sec:discussion}

For any theory that simultaneously provides a solution of the
hierarchy problem and a dark matter particle as a result of a symmetry
guaranteeing its stability, implying pair production of its heavier
partners, the relevant LHC events will be ones in which the heavier
partners are pair produced, with each chain decaying down to largely
visible SM particles and the dark matter particle, which we denote by
$N$.  In many interesting cases, towards the end of each such chain 2
visible SM particles emerge along with the invisible dark matter
particle, {\it e.g.} $Y \to \mu X \to \mu\mu N$, with the preceding
parts of the decay chains giving rise to jets.  In other cases, two
$Y$ particles are directly produced and initiate 2 such chain decays.
In this paper, we have developed a highly effective technique for
using the kinematic information in a typical event containing two $Y
\to \mu X \to \mu\mu N$ decay chains to determine not just the mass
differences in the chain decay, but also the absolute mass scale,
using only the measured $\mu$ momenta and overall visible and missing
transverse momenta.  Since we use purely kinematic information, our
mass determination does not require any assumptions regarding particle
spins, shapes of distributions, cross section and so forth.  
Further, our procedure works whether or not we know 
the topology of each of the chains that
precedes the $Y\to \mu X \to \mu\mu N$ stage. This can be a big
advantage.
For example, in the supersymmetry context 
this allows us to combine $\gl$ and $\wtil q$ initiated chains.

In our study, we have included resolution smearing for muon momenta
and missing momentum as incorporated in the ATLFAST simulation
program. We have also included full combinatorics appropriate to the
$jets+4\mu NN$ final state. Assuming of order 2000 events after cuts
and ATLFAST resolutions appropriate to low-luminosity running,
we have found statistical errors of order $4\gev$ for the individual
$Y$, $X$ and $N$ masses, assuming a reasonable background to signal
ratio, $B/S\lsim 0.5$. There is also a small systematic bias in the
masses extracted.  However, this bias can be removed using Monte Carlo
simulations once the masses are fairly well known. The appropriate
procedure is described in Sec.~\ref{sub:rescomb}.  We have not yet
performed the associated highly computer intensive procedure, but
believe that the systematic biases can be reduced below $1\gev$ (a
residual that we think might arise from possible model dependence of
the kinematic distributions). 

As a particular point of comparison with the many earlier studies that
use the mass-edge technique, we have examined the standard SPS1a
point.  Following our procedure we are left with about 1920 events
(averaging over 10 Monte Carlo ``experiments'') in
the $jets+4\mu$, $jets+2e+2\mu$ and $jets+4e$ channels after cuts
assuming an integrated luminosity of $300\fbi$ and employing
resolutions appropriate to high-luminosity running. The errors on
$\mcntwo$, $\mslep$ and $\mcnone$ are all between $4\gev$ and $5\gev$
if $\wtil \mu$ and $\wtil e$ mass degeneracy is assumed.  The previous
mass-edge studies make this same assumption and employ {\it all} the
final SM particles of the full $\gl\to b \wtil b \to b b \cntwo\to bb
\ell\slep\to bb\ell\ell\cnone$ decay chain but examine only one chain
at a time.  Only one of these mass-edge studies claims an accuracy
($\sim\pm 5\gev$ for $\mcntwo$, $\mslep$ and $\mcnone$) for the same
channels and integrated luminosity that is competitive with the small
error we obtain.

By comparing the SPS1a results obtained for high-luminosity
resolutions to those for this same point using low-luminosity
resolutions (as summarized in the previous section) we found the
important result that the accuracy of our mass determinations was very
little influenced by whether or not we employed low- or
high-luminosity resolution for $\ptmiss$, the latter being essentially
twice the former. That our ability to locate the ``edge'' in a plot of the
number of reconstructed events, $N_{evt}$, as a function of the test
value of, say, $\mcnone$, is not noticeably affected by a factor of
two deterioration in resolution for $\ptmiss$ is a sign of the
robustness of our approach. 

Accuracies of order $4-5\gev$ for the masses of new-physics particles
will yield an accurate determination of the TeV-scale underlying
parameters of the associated new physics model. The latter accuracy
will, in turn, typically yield reasonably precise evolved values for
the model parameters at any higher scale ({\it e.g.}  the coupling
constant unification scale in SUSY) where they might follow a
meaningful pattern that would determine the more fundamental structure
of the new physics theory.  Further, an accuracy of order $4-5\gev$
for the dark matter particle mass will in many cases allow a
sufficiently accurate calculation for the dark matter density from
annihilations in the early universe as to allow a meaningful
comparison with the very accurate observed value.  In some cases, the
dark matter particle coannihilates with another particle of only
slightly larger mass.  We will be exploring the extent to which the
mass of the coannihilation partner could be determined in such a
situation. For the moment, we can only claim the $4-5\gev$ kind of
error on individual masses when mass differences are reasonably
substantial (and the number of events after cuts is of order 1700 to
2000).

A `fun' example that we hope our experimental colleagues will
pursue is to employ our method for determining the mass scales for the
top and $W$ simultaneously in the $t\bar{t}$ di-lepton decay topology.
Or, given that the $W$ mass is already quite well-known, they could
impose this additional constraint in our context and get an excellent
$t$ mass determination.

The heart of our technique is the fact that by considering both decay
chains in a typical LHC event together, a choice for the chain decay
masses $\calm=\{m_Y,m_X,m_N\}$ (see Fig.~\ref{fig:topology}) in
combination with the measured momenta of the 4 visible and measurable
SM particles emitted in the two chains implies a discrete (sometimes
even unique) set of three momenta for the two final state $N$'s. (One
is solving a quartic equation.)  Conversely, if we have already used
our procedure to determine to good precision the
$\calm=\{m_Y,m_X,m_N\}$ masses, we can invert the process.  For each
event, we can input the known masses and obtain a set of discrete
choices for the momenta, $\vec p_N$ and $\vec p_{N'}$, of the final
invisible particles.  For each discrete choice, the 4-momenta of all
particles in the decay chains are then determined.  These 4-momenta
can then be input to a given model (with definite spins for the
$Y,X,N$ and definite decay correlations and so forth).  One can then
test the experimental distributions ({\it e.g.} of correlation angles,
of masses constructed from the visible SM particles, and so forth)
against predictions obtained for the model using a Monte Carlo.
Presumably, this will provide strong discrimination between different
models that have the same already-determined chain decay masses.  The
only question is to what extent the possibility of more than one
discrete solution for each event will confuse the distributions
obtained from the Monte Carlo.

Conversely, it is clear that determining the spins of all the
particles in a chain of decays can be difficult without a relatively
precise {\it purely-kinematic} determination of the masses .  In
particular, we expect that angular correlations and the like (obtained
from Monte Carlos that assume a particular model including spins) will
be strongly influenced by the masses. Confusion between two different
models with differing spins and masses can be anticipated in the
absence of an independent purely-kinematical determination
of the masses.

Overall, we claim that our techniques provide some powerful new tools
for doing precision physics at the LHC in an environment where new
physics events contain invisible particles of unknown mass.  We hope
the experimental community will pursue the approaches we have
analyzed. We do not anticipate that fully realistic simulations will
lead to significantly larger errors for new particle masses than those
we have found, but it is clearly important to verify that this is the
case.

\acknowledgments

\vspace*{-.1in}
This work was supported in part by U.S. Department of Energy grant No. DE-FG03-91ER40674.
JFG and HCC thank the Aspen Center for Physics where a portion of this work
was performed. 

\appendix

\flushleft{\bf \large Appendices}

\section{Solution Procedure}
\label{sec:relations}
To determine whether a given event with the topology of Fig.~\ref{fig:topology} 
is consistent with a given mass hypothesis, we proceed as follows.
We envision the process of $pp\to (135)+(246)$, followed by
$(135)\to 5+(31)$ and $(246)\to 6+(42)$, which in turn is followed by
$(31)\to 3+1$ and $(42)\to 4+2$.  (The objects in $(\ldots)$ are to
be thought of as single on-shell particles: in the notation of
Fig.~\ref{fig:topology}, $(135)=Y$, $(246)=Y'$, $(13)=X$, $(24)=X'$,
$1=N$ and $2=N'$.) We will be assuming input values for
$m_{135}^2$, $m_{246}^2$, $m_{13}^2$, $m_{24}^2$, $m_1^2$ and $m_2^2$,
assuming $m_{135}^2=m_{246}^2$, $m_{13}^2=m_{24}^2$ and $m_1^2=m_2^2$.
The cross section takes the form
\bea
d\sig&=&{1\over 2s(2\pi)^8}\int dx_1 dx_2 |\calm|^2 f(x_1)f(x_2) 
 {d^3\vec p_5\over 2E_5}{d^3\vec p_6\over 2E_6}{d^3\vec p_3\over 2E_3}{d^3\vec p_4\over
   2E_4}{d^3\vec p_1\over 2E_1}{d^3\vec p_2\over 2E_2}\nn\\
&&\qquad\qquad\times \del^4\left[x_1p_A+x_2p_B
 -(p_1+p_2+p_3+p_4+p_5+p_6)\right]  
\eea
We first convert 
\beq
dx_1dx_2={2\over s}dE_{tot}dp^z_{tot}\,,
\eeq
introduce on-shell masses for the intermediate particles, 
and introduce on-shell $\del$ functions for the invisible particles
$1$ and $2$ to yield
\bea
d\sig&=&{1\over 4(2\pi)^8}\int dE_{tot}dp^z_{tot} dm_{135}^2dm_{246}^2
dm_{31}^2 dm_{24}^2|\calm|^2 f(x_1)f(x_2) 
 {d^3\vec p_5\over 2E_5}{d^3\vec p_6\over 2E_6}{d^3\vec p_3\over 2E_3}{d^3\vec p_4\over
   2E_4}\nn\\
&&\qquad\qquad\times {d^4 p_1}\del(p_1^2-m_1^2){d^4 p_2}\del(p_2^2-m_2^2)\nn\\
&&\qquad\qquad\times \del^4\left[x_1p_A+x_2p_B
 -(p_1+p_2+p_3+p_4+p_5+p_6)\right]\nn\\
&&\qquad\qquad \times \del[(p_1+p_3+p_5)^2-m_{135}^2]\del[
(p_2+p_4+p_6)^2  -m_{246}^2]\nn\\
&&\qquad\qquad\times
\del[(p_3+p_1)^2-m_{31}^2]\del[(p_2+p_4)^2-m_{42}^2]\nn\\
&=&{1\over 4(2\pi)^8}\int dp^z_{tot} dm_{135}^2dm_{246}^2
dm_{31}^2 dm_{24}^2|\calm|^2 f(x_1)f(x_2) 
 {d^3\vec p_5\over 2E_5}{d^3\vec p_6\over 2E_6}{d^3\vec p_3\over 2E_3}{d^3\vec p_4\over
   2E_4}\nn\\
&&\qquad\qquad\times {d^4 p_1}\del(p_1^2-m_1^2){d^4 p_2}\del(p_2^2-m_2^2)\nn\\
&&\qquad\qquad\times  \del[(p_1+p_3+p_5)^2-m_{135}^2]\del[
(p_2+p_4+p_6)^2  -m_{246}^2]\nn\\
&&\qquad\qquad\times
\del[(p_3+p_1)^2-m_{31}^2]\del[(p_2+p_4)^2-m_{42}^2]\nn\\
\eea
where in the last step we eliminated $d^3\vec p_2$ using the
3-momentum conservation part of the $\del^4$ function and eliminated
$E_{tot}$ using the energy part of the $\del^4$ function.
For fixed values of the unknown masses,
we end up with the 4 unknowns of $p_{tot}^z$ and $\vec p_1$,
to be solved for using the 4 on-shell $\del$ functions. 
We will now define 
\beq
p_{vis}\equiv p_3+p_5+p_4+p_6\,.
\eeq
Assuming no transverse momentum for the  $Y+Y'=1+2+3+4+5+6$ system, we
then have
\bea
p_1\cdot p_3&=&E_1E_3-p_1^zp_3^z-p_1^yp_3^y-p_1^x p_3^x\nn\\
p_2\cdot p_4&=&E_2E_4-(p_{tot}^z-p_{vis}^z-p_1^z)p_4^z-(-p_{vis}^y-p_1^y)p_4^y-(-p_{vis}^x-p_1^x) p_4^x\nn\\
p_1\cdot p_5&=&E_1E_5-p_1^zp_5^z-p_1^yp_5^y-p_1^x p_5^x\nn\\
p_2\cdot p_6&=&E_2E_6-(p_{tot}^z-p_{vis}^z-p_1^z)p_6^z-(-p_{vis}^y-p_1^y)p_6^y-(-p_{vis}^x-p_1^x) p_6^x\,.
\eea
(Transverse momentum for the $Y+Y'$ system can, and must, be included in the
obvious way. We compute it as the negative of the sum
of the observed momenta of particles 3, 4, 5 and 6 and the missing momentum.)
We next combine the last two $\del$ functions and consider the
requirement (again, recall that we are assuming some input mass values for
the intermediate on-shell particle masses)
\beq
2p_3\cdot p_1-2 p_2\cdot p_4+\Delta_{2b}\equiv G_1=0\,,
\eeq
where 
\beq
\Delta_{2b}\equiv m_3^2+m_1^2-m_{31}^2+m_{42}^2-m_2^2-m_4^2\,.
\eeq
Similarly we combine the $135$ and $246$ $\del$ functions to obtain
\beq
2p_1\cdot p_5-2p_2\cdot p_6+\Delta_{3b}\equiv G_2=0\,,
\eeq
where 
\beq
\Delta_{3b}\equiv
m_{31}^2+m_5^2-m_{531}^2+m_{642}^2-m_{42}^2-m_6^2
+2p_3\cdot p_5-2p_4\cdot p_6\,.
\eeq
Of course, $m_5^2$ and $m_6^2$ are measured
experimentally (and are typically small unless one is a $W$ or $Z$), 
and $2p_3\cdot p_5$ and $2p_4\cdot p_6$ are also computable from the
experimental event. Further, we are assuming input values for $m_{31}^2$,
$m_{42}^2$, $m_{531}^2$ and $m_{642}^2$.  The above is a convenient
organization, since $\Delta_{2b}=0$ and
$\Delta_{3b}$ reduces to just  momenta dot products when the
two decay chains are identical. 

We now implement directly the $m_{31}^2$ and $m_{531}^2$ $\del$
functions.
\beq
m_{31}^2-m_1^2-m_3^2-2p_1\cdot p_3\equiv \Delta_{31}^2-2p_1\cdot
p_3\equiv G_3=0\,,
\eeq
and
\beq
m_{531}^2-m_{31}^2-m_5^2-2p_3\cdot p_5-2p_1\cdot
p_5\equiv\Delta_{531}-2p_1\cdot p_5\equiv G_4=0\,,
\eeq
where again $2p_3\cdot p_5$ is determined experimentally and the
masses are being input.

We now solve these 4 equations for the 4 unknowns of $p_{tot}^z$,
$p_1^z$, $p_1^y$, and $p_1^x$.  We write the solutions in the form:
\bea
p_1^x&=&c_{xe1}E_1+c_{xe2}E_2+c_x\nn\\
p_1^y&=&c_{ye1}E_1+c_{ye2}E_2+c_y\nn\\
p_1^z&=&c_{ze1}E_1+c_{ze2}E_2+c_z\nn\\
p_{tot}^z&=& c_{zte1}E_1+c_{zte2}E_2+c_{zt}
\label{p1forms}
\eea
where the $c$'s above are somewhat complicated 
functions of the masses, energies and momenta of the
visible particles, 3, 4, 5, and 6.
The Jacobian for the variable change $p_1^x,p_1^y,p_1^z,p_{tot}^z\to G_1,G_2,G_3,G_3$
 is easily computed as
\bea
J&=& 16\bigl[ - p_3^z p_4^z p_5^y p_6^x +  p_3^y p_4^z p_5^z p_6^x  
     +   p_3^z p_4^z p_5^x p_6^y -  p_3^x p_4^z p_5^z p_6^y\nn\\
&&\qquad\qquad 
     -   p_3^z p_4^y p_5^x p_6^z +  p_3^z p_4^x p_5^y p_6^z  
     -   p_3^y p_4^x p_5^z p_6^z +  p_3^x p_4^y p_5^z p_6^z\bigr]
\eea
It is a function only of observed momenta.
For the next stage, we combine the expressions of Eq.~(\ref{p1forms})
with
\bea
p_2^x = -p_{vis}^x - p_1^x\,,\quad
p_2^y = -p_{vis}^y-p_1^y\,,\quad
p_2^z = p_{tot}^z-p_{vis}^z-p_1^z\,,
\label{p2forms}
\eea
and solve the equations for the on-shell $\del$ functions
for $p_1$ and $p_2$ (the invisible particles)
\bea
0&=& E_1^2 - (p_1^x)^2 - (p_1^y)^2 - (p_1^z)^2-m_1^2\label{eqaf}\\
0 &=& E_2^2 - (p_2^x)^2 - (p_2^y)^2 - (p_2^z)^2-m_2^2\label{eqbf}
\eea
for $E_1$ and $E_2$. For convenience, we rewrite 
Eqs.~(\ref{eqaf}) and (\ref{eqbf}) in the respective forms:
\bea
&&a_{11}E_1^2+a_{12}E_1E_2+a_{22}E_2^2+a_1 E_1+a_2 E_2+a\equiv F_A=0\label{eqa}\\
&&b_{11}E_1^2+b_{12}E_1E_2+b_{22}E_2^2+b_1 E_1+b_2 E_2+b\equiv F_B=0\label{eqb}
\eea
where the $a_{ij}$, $b_{ij}$, $a_i$, $b_i$ as well as $a$ and $b$ are
functions of the $c$'s, $m_1^2$, $m_2^2$ and the components of $\vec p_{vis}$. 
We now take
\beq
F_A - {a_{11}\over b_{11}} \times F_B=0
\eeq
and solve the resulting linear equation for $E_1$ to obtain
\bea
E_1&=& { a_{11}\ b - a\ b_{11} - a_2\ b_{11}\ E_2 + a_{11}\ b_2\ e_2 - a_{22}\ b_{11}\ E_2^2 + \
a_{11}\ b_{22}\ E_2^2 \over
 -a_{11} b_1 + a_1\ b_{11} + a_{12}\ b_{11}\ E_2 - a_{11}\ b_{12}\
 E_2}
\label{e1lin}
\eea
We now substitute this result into Eq.~(\ref{eqa}) to obtain the final
quartic equation for $E_2$ of form
\beq
A E_2^4+B E_2^3 +C E_2^2 +D E_2 +E=0\,,
\eeq
where $A$, $B$, $C$, $D$ and $E$ are functions of the $a_{ij}$,
$b_{ij}$, $a_i$, $b_i$, $a$ and $b$.
We then employ a standard computer subroutine for obtaining the 4
roots of this quartic equation. For typical input visible momenta,
some of the roots will be acceptable real solutions and some will be
imaginary. We retain all real solutions.
(The Jacobian for the $F_A,F_B\to E_1,E_2$ transformation is easily
computed.)
Once real values for $E_1$ and $E_2$ are obtained these can be
substituted into Eq.~(\ref{p1forms}) to determine the 3-vector components
of $p_1$ and the $z$ component of $p_{tot}$.  The components of $p_2$
are then obtained by momentum conservation.  At this point, the
invisible 4-momenta are fully determined and could potentially be
employed in a model matrix element.

\section{SUSY points}
\label{sec:points}

In this appendix, we give details regarding the SUSY points simulated.

\subsection{Point I}

We input low-scale parameters of
\bea
&\mu=+300\gev,\quad \tanb=10,\quad
(\wt M_1,\wt M_2,\wt M_3)=(90,300,500)\gev&
\nonumber\\
&\wt m_L^{(1,2,3)}=\wt m_E^{(3)}=1000\gev,\quad \wt m_E^{(1,2)}=120\gev&
\nonumber\\
&\wt m_Q^{(1,2)}=400\gev,\quad \wt
m_{U,D}^{(1,2)}=300\gev,\quad \wt m_Q^{(3)}= \wt m_{U,D}^{(3)}=1000\gev&
\eea
where the $\wt m$'s are the soft slepton and squark masses, and the 
$\wt M$'s are the gaugino masses. $L$ and $Q$ refer to the slepton and
squark $SU(2)_W$ doublets and $E$, $U$ and $D$ refer to the slepton and squark
singlets. Superscripts give the generations. 
The decay chain of interest is 
\beq
\wt q_L\to q\cntwo\quad \cntwo\to \mu
\wt \mu_R,\quad \wt \mu_R\to \mu \cnone\,.
\eeq
 Using the input soft parameters
as specified above and SPheno~2.2.3~\cite{Porod:2003um}, the sparticle
masses of relevance for our discussion are (all in GeV):
\bea
&\mgl\sim524,\quad  m_{\wt d_L,\wt s_L}\sim 438,\quad  m_{\wt u_L,\wt
  c_L}\sim 431&\nonumber\\
&\mcntwo\sim 246.6,\quad m_{\wt \mu_R}\sim
128.4,\quad \mcnone\sim 85.3&
\eea
For this point, the net cross section available is 
\beq
\sigma\left(pp\to \sum_{q,q'=u,d,c,s} \wt q_L\,\wt q_L'+ \sum_{q,q'=u,d,c,s}\wt q_L\,\anti{\wt q_L'}+ \sum_{q,q'=u,d,c,s}\anti{\wt q_L}\,\anti{\wt q_L'}\right)\sim 2.9\times 10^4\fb\,,
\eeq
coming from all sources including $gg$ fusion, $u_Lu_L$ fusion, {\it etc.}
The branching ratios relevant to the particular decay chain we examine
are
\bea
&&\br(\wt q_L\to q\cntwo )\sim 0.27\quad(q=u,d,c,s)
\nonumber\\
&&\br(\cntwo\to \wt\mu_R^\pm \mu^\mp)\sim 0.124
\nonumber\\
&&\br(\wt\mu_R^\pm\to\mu^\pm \cnone)= 1\,.
\eea
The net effective branching ratio for the double decay chain is
\beq
\br(\wt q_L\wt q_L\to 4\mu \cnone\cnone)\sim (0.27)^2\times (0.124)^2\sim 1.12\times 10^{-3}
\eeq
for any one $\wt q_L$ choice.  The effective cross section for the
$4\mu\cnone\cnone$ final state is then
\beq
\sigma(4\mu\cnone\cnone)\sim 2.9\times 10^4\fb\times 1.12\times
10^{-3}\sim 32.5 \fb\,.
\eeq
For an integrated luminosity of $L=90\fbi$, this gives us 2900
$4\mu\cnone\cnone$ events before any cuts are applied. After cuts, we
are left with about 1900 events.

\subsection{Point II}

Point II is defined by the following input low-scale  SUSY parameters:
\bea
&\mu=+300\gev,\quad \tanb=10,\quad
(\wt M_1,\wt M_2,\wt M_3)=(90,200,500)\gev&
\nonumber\\
&\wt m_L^{(1,2,3)}=\wt m_E^{(3)}=1000\gev,\quad \wt m_E^{(1,2)}=140\gev&
\nonumber\\
&\wt m_Q^{(1,2)}=400\gev,\quad \wt
m_{U,D}^{(1,2)}=300\gev,\quad \wt m_Q^{(3)}= \wt m_{U,D}^{(3)}=1000\gev\,.&
\eea
Using SPheno~2.2.3~\cite{Porod:2003um}, the relevant chain-decay masses are
\beq
\{m_Y=\mcntwo,m_X=m_{\wt\mu_R},m_N=\mcnone\}=\{180.8,147.1,85.2\}\gev\,.
\eeq
However, we do not employ the cross sections and branching ratios
predicted by these parameters.  Instead, we assume 2000
available experimental points after cuts, close to the 1900 left after cuts in
the case of Point I.  This allows us to see
how errors change in the case where $m_Y-m_X$ is much smaller than in
the case of Point I.
\subsection{Point III}

The masses used in this case are obtained from PYTHIA
1.0.8~\cite{Sjostrand:2006za}
using the low-scale parameters
\bea
&\mu=+3000\gev,\quad \tanb=10,\quad
(\wt M_1,\wt M_2,\wt M_3)=(0.2,200,500)\gev&
\nonumber\\
&\wt m_L^{(1,2,3)}=\wt m_E^{(3)}=1000\gev,\quad \wt m_E^{(1,2)}=100\gev&
\nonumber\\
&\wt m_Q^{(1,2)}=400\gev,\quad  \wt
m_{U,D}^{(1,2)}=300\gev,\quad \wt m_Q^{(3)}=\wt m_{U,D}^{(3)}=1000\gev&
\eea
yielding  
\beq
\{m_Y=\mcntwo,m_X=m_{\wt\mu_R},m_N=\mcnone\}=\{199.4,100.0,0.1\}\gev\,.
\eeq
Again, we assume
2000 available experimental points after cuts, close to the 1900 
events after cuts obtained for Point I.

\subsection{Point IV: SPS1a}

For the SPS1a point, we use the GUT-scale
mSUGRA inputs of
\beq
m_{1/2}=250\gev, \quad m_0=100\gev,\quad A_0=-100\gev,\quad \mbox{tan}\beta=10,\quad \mu>0.
\eeq
From ISAJET 7.75, the spectrum is calculated as 
\bea
&\mgl\sim 608\gev,\quad  m_{\wt d_L,\wt s_L}\sim 571\gev,\quad  m_{\wt u_L,\wt
  c_L}\sim 565\gev&\nonumber\\
&\mcntwo\sim 180.3\gev,\quad m_{\wt \mu_R}\sim
142.5\gev,\quad \mcnone\sim 97.4\gev\quad\mstauone\sim 134.7\gev&\nonumber\\
\eea
The total effective cross-section including all channels for
$\cntwo\cntwo$ production is about $1\pb$. The relevant branching ratios are:
\beq
\br(\cntwo\to \wt\mu_R^\pm \mu^\mp)\sim 0.063,
\quad
\br(\wt\mu_R^\pm\to\mu^\pm \cnone)= 1\,.
\eeq
For $L=300\fbi$, this gives 1200 events before any cuts. After cuts,
we are left with about 425 events.  Errors for the masses given in the text
are based upon the latter.


\begin{thebibliography}{99}

\bibitem{ued}
  T.~Appelquist, H.~C.~Cheng and B.~A.~Dobrescu,
  ``Bounds on universal extra dimensions,''
  Phys.\ Rev.\  D {\bf 64}, 035002 (2001)
  [arXiv:hep-ph/0012100].


\bibitem{Cheng:2001an}
  H.~C.~Cheng, D.~E.~Kaplan, M.~Schmaltz and W.~Skiba,
  ``Deconstructing gaugino mediation,''
  Phys.\ Lett.\  B {\bf 515}, 395 (2001)
  [arXiv:hep-ph/0106098];
  ``Bosonic supersymmetry? Getting fooled at the LHC,''
  Phys.\ Rev.\  D {\bf 66}, 056006 (2002)
  [arXiv:hep-ph/0205314].
  
\bibitem{lht}
  H.~C.~Cheng and I.~Low,
  ``TeV symmetry and the little hierarchy problem,''
  JHEP {\bf 0309}, 051 (2003)
  [arXiv:hep-ph/0308199];
  ``Little hierarchy, little Higgses, and a little symmetry,''
  JHEP {\bf 0408}, 061 (2004)
  [arXiv:hep-ph/0405243].

\bibitem{Agashe:2004ci}
  K.~Agashe and G.~Servant,
  ``Warped unification, proton stability and dark matter,''
  Phys.\ Rev.\ Lett.\  {\bf 93}, 231805 (2004)
  [arXiv:hep-ph/0403143].

\bibitem{Baltz:2006fm}
  E.~A.~Baltz, M.~Battaglia, M.~E.~Peskin and T.~Wizansky,
  Phys.\ Rev.\  D {\bf 74}, 103521 (2006)
  [arXiv:hep-ph/0602187].
  
\bibitem{Hinchliffe:1996iu}
  I.~Hinchliffe, F.~E.~Paige, M.~D.~Shapiro, J.~Soderqvist and W.~Yao,
  ``Precision SUSY measurements at LHC,''
  Phys.\ Rev.\  D {\bf 55}, 5520 (1997)
  [arXiv:hep-ph/9610544].


\bibitem{cambridge}
  C.~G.~Lester and D.~J.~Summers,
  ``Measuring masses of semi-invisibly decaying particles pair produced at
  hadron colliders,''
  Phys.\ Lett.\  B {\bf 463}, 99 (1999)
  [arXiv:hep-ph/9906349];
  A.~Barr, C.~Lester and P.~Stephens,
  ``m(T2): The truth behind the glamour,''
  J.\ Phys.\ G {\bf 29}, 2343 (2003)
  [arXiv:hep-ph/0304226].
  
\bibitem{Bachacou:1999zb}
  H.~Bachacou, I.~Hinchliffe and F.~E.~Paige,
  ``Measurements of masses in SUGRA models at LHC,''
  Phys.\ Rev.\  D {\bf 62}, 015009 (2000)
  [arXiv:hep-ph/9907518].

\bibitem{Allanach:2000kt}
  B.~C.~Allanach, C.~G.~Lester, M.~A.~Parker and B.~R.~Webber,
  ``Measuring sparticle masses in non-universal string inspired models at  the
  LHC,''
  JHEP {\bf 0009}, 004 (2000)
  [arXiv:hep-ph/0007009]. See also C.~G.~Lester, {\tt
    http://cdsweb.cern.ch/search.py?sysno=002420651CER}.

\bibitem{Gjelsten:2004ki}
  B.~K.~Gjelsten, D.~J.~Miller and P.~Osland,
  ``Measurement of SUSY masses via cascade decays for SPS1a,''
  JHEP {\bf 0412}, 003 (2004)
  [arXiv:hep-ph/0410303].

\bibitem{Weiglein:2004hn}
  G.~Weiglein {\it et al.}  [LHC/LC Study Group],
  Phys.\ Rept.\  {\bf 426}, 47 (2006)
  [arXiv:hep-ph/0410364], Sections 5.1 and 5.2.


\bibitem{Kawagoe:2004rz}
  K.~Kawagoe, M.~M.~Nojiri and G.~Polesello,
  ``A new SUSY mass reconstruction method at the CERN LHC,''
  Phys.\ Rev.\  D {\bf 71}, 035008 (2005)
  [arXiv:hep-ph/0410160].

\bibitem{Allanach:2004ub}
  B.~C.~Allanach {\it et al.}  [Beyond the Standard Model Working Group],
  arXiv:hep-ph/0402295; see the contribution by C.~G.~Lester, Section X.



\bibitem{Lester:2005je}
  C.~G.~Lester, M.~A.~Parker and M.~J.~White,
  ``Determining SUSY model parameters and masses at the LHC using
  cross-sections, kinematic edges and other observables,''
  JHEP {\bf 0601}, 080 (2006)
  [arXiv:hep-ph/0508143].



\bibitem{Arkani-Hamed:2005px}
  N.~Arkani-Hamed, G.~L.~Kane, J.~Thaler and L.~T.~Wang,
  ``Supersymmetry and the LHC inverse problem,''
  JHEP {\bf 0608}, 070 (2006)
  [arXiv:hep-ph/0512190].

\bibitem{Butterworth:2007ke}
  J.~M.~Butterworth, J.~R.~Ellis and A.~R.~Raklev,
  ``Reconstructing sparticle mass spectra using hadronic decays,''
  arXiv:hep-ph/0702150.

\bibitem{Allanach:2002nj}
  B.~C.~Allanach {\it et al.},
  ``The Snowmass points and slopes: Benchmarks for SUSY searches,''
in {\it Proc. of the APS/DPF/DPB Summer Study on the Future of Particle Physics (Snowmass 2001) } ed. N.~Graf,
{\it In the Proceedings of APS / DPF / DPB Summer Study on the Future of Particle Physics (Snowmass 2001), Snowmass, Colorado, 30 Jun - 21 Jul
2001, pp P125}
  [arXiv:hep-ph/0202233].

\bibitem{Miller:2005zp}
  D.~J.~Miller, P.~Osland and A.~R.~Raklev,
  ``Invariant mass distributions in cascade decays,''
  JHEP {\bf 0603}, 034 (2006)
  [arXiv:hep-ph/0510356].

\bibitem{dirk}
We particularly thank Dirk Zerwas for several detailed discussions and emails.

\bibitem{Gleisberg:2003xi}
  T.~Gleisberg, S.~Hoche, F.~Krauss, A.~Schalicke, S.~Schumann and J.~C.~Winter,
  JHEP {\bf 0402}, 056 (2004)
  [arXiv:hep-ph/0311263].

\bibitem{Sjostrand:2006za}
  T.~Sjostrand, S.~Mrenna and P.~Skands,
  JHEP {\bf 0605}, 026 (2006)
  [arXiv:hep-ph/0603175].

\bibitem{Porod:2003um}
  W.~Porod,
  Comput.\ Phys.\ Commun.\  {\bf 153}, 275 (2003)
  [arXiv:hep-ph/0301101].

\bibitem{atlfast}
  E. Richter-Was, D. Froidevaux and L. Poggioli, ``ATLFAST 2.0: a fast
  simulation package for ATLAS'', Tech. Rep. ATL-PHYS-98-131 (1998);
  see also,   H.T. Phillips, P.Clarke, E.Richter-Was, P.Sherwood, R.Steward
  [{\tt http://root.cern.ch/root/Atlfast.html}].

\end{thebibliography}
\end{document}